\documentclass[useAMS,usenatbib]{mn2e}
\usepackage{amsmath}
\usepackage{graphicx}
\usepackage{float}
\usepackage{anysize}
\usepackage{times}

\usepackage[usenames,dvipsnames,table]{xcolor}
\title[Constraints from 2M++]{ 
Cosmological parameters from the
comparison of peculiar velocities with predictions from the 2M++ density field}

\author[J. Carrick et al.]
  {Jonathan Carrick,$^1$
  Stephen J.~Turnbull,$^1$
  Guilhem Lavaux,$^{1,2,3,4,5}$
  \newauthor
  Michael J. Hudson$^{1,4}$\\
  $^1$Department of Physics \& Astronomy, University of Waterloo, Waterloo, ON,
  N2L 3G1, Canada\\
  $^2$CNRS, UMR7095, Institut d'Astrophysique de Paris, F-75014, Paris, France\\
  $^3$Sorbonne Universit\'{e}s, UPMC Univ Paris 06, UMR7095, Institut d'Astrophysique de Paris, F-75014, Paris, France\\  
  $^4$Perimeter Institute for Theoretical Physics, 31 Caroline St. N., Waterloo,
  ON, N2L 2Y5, Canada\\
  $^5$Canadian Institute for Theoretical Astrophysics, University of Toronto,
  60 St. George Street, Toronto, ON M5S 1A7, Canada\\
  email: mjhudson@uwaterloo.ca }

\date{\today} 

\pagerange{000--000} \pubyear{0000}

\def \b{$\beta^*$}
\def \fsig{$f\sigma_{8}$}
\def \fslin{$f\sigma\sbr{8,lin}$}
\def \OmSig{$\Omega_{\rm{m}}^{0.55}\sigma\sbr{8}$}
\def \Vext{\mamo{\boldmath{V}\sbr{ext}}}

\def \bestb{0.431 $\pm$ 0.021}
\def \bestfsig{0.401 $\pm$ 0.024}

\newcommand{\apj}{ApJ}
\newcommand{\apjl}{ApJL}
\newcommand{\apjs}{ApJS}
\newcommand{\mnras}{MNRAS}
\newcommand{\aj}{AJ}

\newcommand{\jcap}{JCAP}

\newcommand{\nat}{Nature}

\def\gtwid{\mathrel{\raise.3ex\hbox{$>$\kern-.75em\lower1ex\hbox{$\sim$}}}}
\def\ltwid{\mathrel{\raise.3ex\hbox{$<$\kern-.75em\lower1ex\hbox{$\sim$}}}}
\def\\{\hfil\break}
\def\ie{{\it i.e.\ }}
\def\eg{{\it e.g.\ }}

\def\lesssim{\mathrel{\hbox{\rlap{\hbox{\lower4pt\hbox{$\sim$}}}\hbox{$<$}}}}
\def\gtrsim{\mathrel{\hbox{\rlap{\hbox{\lower4pt\hbox{$\sim$}}}\hbox{$>$}}}}

\def\arcdeg{\hbox{$^\circ$}}

\newcommand{\mamo}[1]{\mbox{$#1$}}
\newcommand{\unit}[1]{\ifmmode \:\mbox{\rm #1}\else \mbox{#1}\fi}

\newcommand{\sbr}[1]{_{\rm #1}}

\newcommand{\mone}{\mamo{^{-1}}}

\newcommand{\kms}{\unit{km~s\mone}}

\newcommand{\mpc}{\unit{Mpc}}

\newcommand{\hmpc}{\mamo{h\mone}\mpc}

\newcommand{\lberr}[4]{\mamo{l = #1\arcdeg \pm #2\arcdeg}, \mamo{b = #3\arcdeg \pm #4\arcdeg}}

\newcommand{\wrt}{with respect to}

\newcommand{\secref}[1]{Section~\ref{sec:#1}}

\newcommand{\figref}[1]{Fig.~\ref{fig:#1}}

\begin{document}

\label{firstpage}

\maketitle

\begin{abstract}
  Peculiar velocity measurements are the only tool available in the
  low-redshift Universe for mapping the large-scale distribution of
  matter and can thus be used to constrain cosmology. Using redshifts
  from the 2M++ redshift compilation, we reconstruct the density of
  galaxies within 200 \hmpc, allowing for the first time good sampling
  of important superclusters such as the Shapley Concentration. We
  compare the predicted peculiar velocities from 2M++ to Tully-Fisher and SNe
  peculiar velocities. We find a value of \b\ $\equiv
  \Omega_{\rm{m}}^{0.55}/b^* =$ \bestb, suggesting
  $\Omega_{\rm{m}}^{0.55}\sigma_{\rm{8,lin}}$ $=$ \bestfsig,
  in good agreement with other probes. 
  The predicted peculiar velocity of the Local Group arising 
  from the 2M++ volume alone is $540 \pm 40$
  \kms, towards \lberr{268}{4}{38}{6}, only $10^\circ$ out of
  alignment with the Cosmic Microwave Background dipole.  To account for velocity
  contributions arising from sources outside the 2M++ volume, we fit
  simultaneously for \b\ and an external bulk flow in our analysis.
  We find that an external bulk flow is preferred at the 5.1$\sigma$
  level, and the best fit has a velocity of $159\pm23 \kms$ towards
  \lberr{304}{11}{6}{13}.  Finally, the predicted bulk flow of a 50
  \hmpc\ Gaussian-weighted volume centred on the Local Group is 230
  $\pm$ 30 \kms, in the direction \lberr{293}{8}{14}{10}, in agreement
  with predictions from $\Lambda$CDM.
\end{abstract}

\section{Introduction}
Peculiar velocities, i.e.\ deviations in the motions of galaxies from
the Hubble flow, are valuable tools which probe the underlying
distribution of dark matter, and are in fact the only practical means
of doing so on large scales in the low redshift Universe.  They can be
used to constrain the amplitude of matter density fluctuations on a
range of scales. As the amplitude of such fluctuations are themselves
cosmology-dependent, the analysis of peculiar velocities provides a
direct means of testing cosmological predictions.

In the current standard cosmological paradigm, the observed structure
in the Universe is a result of gravitational instabilities which grew
from density perturbations in an otherwise homogeneous
background. This gravitational attraction of objects to surrounding
structure results in peculiar motion, \ie motion in addition to that
resulting from the expansion of the Universe. Assuming only mass
continuity and standard gravitation in an expanding universe, in the
linear regime where these fluctuations in density are small, \ie
$\delta \mathbf(r) = (\rho - \bar{\rho})/\bar{\rho} \lesssim 1$,
peculiar velocities are proportional to gravitational
accelerations. This relation, as expressed in integral form, is as
follows:
\begin{equation}
\mathbf{v}(\mathbf{r}) = \frac{f(\Omega_{\rm m})}{4\pi}\int
d^3\mathbf{r}'\delta(\mathbf{r}')\frac{(\mathbf{r}' - \mathbf{r})}{|\mathbf{r}'
- \mathbf{r}|^3}\; ,
\label{main}
\end{equation}
\noindent where $\mathbf{v}(\mathbf{r})$ is the peculiar velocity
field, $\delta(\mathbf{r})$ is the mass density contrast, $\Omega_{\rm
  m}$ is the cosmological density parameter, and where distances are
measured in \kms\ (\ie $r = HR$, where $H$ is the Hubble parameter and
$R$ is the comoving distance in Mpc). The growth rate of density
perturbations, $f(\Omega_{\rm m})$, is generally parameterized by
$\Omega_{\rm{m}}^\gamma$, where $\gamma=0.55$ for $\Lambda$CDM
\citep{Wang98}, which has recently been shown to be consistent with
observations \citep{HudTur12}.

As the total matter density contrast cannot be observed, however, to
make use of Equation (\ref{main}), an assumption must first be made as
to how observed galaxies trace the underlying total matter. Assuming
linear biasing holds on large scales, $\delta_{\rm{g}} = b\delta$,
where $b$ is the linear bias factor, and where we have used the
subscript ``$\rm{g}$'' when referring to galaxies. Rewriting Equation
(\ref{main}) in terms of the observable density contrast of galaxies
in the nearby Universe, the proportionality factor between
gravitational acceleration and peculiar velocity is then $\beta \equiv
f/b$.  Thus, if linear theory holds, by comparing measured peculiar
velocities to those predicted by the distribution of galaxies in
redshift surveys, one can constrain cosmological parameters through
the measurement of $\beta$.  Furthermore, under the assumption of
linear biasing $\sigma_{\rm{8,g}} = b\sigma_{\rm{8}}$, where
$\sigma_{\rm{8}}$ is the root mean square density fluctuations on an 8
\hmpc\ scale.

By measuring $\sigma_{\rm{8,g}}$ directly from the redshift data, one
can eliminate $b$ and constrain the degenerate cosmological parameter
combination $f\sigma_{\rm{8}} = \beta \sigma_{\rm{8,g}}$. Using this
method of velocity-velocity comparison, a number of recent studies
have constrained this parameter combination
\citep{PikHud05, DavNusMas11, BraDavNus12, TurHudFel12}.

In addition to constraints placed on these parameters, when full sky
surveys are used one can compute the velocity of the Local Group
(hereafter LG) arising from the volume under consideration as
predicted by linear theory. While it has long been assumed that the 
dipole in the Cosmic Microwave Background (CMB) temperature map 
is a Doppler effect due to the Sun's motion, this has only been proved recently,
via the aberration of the CMB temperature anisotropies \citep{Planck13-27}. 
The motion of the Sun with respect to the Galaxy and of the Galaxy 
with respect to the barycentre of the LG are well known 
(e.g.\ \citealt{Couvan99}), and when combined with 
the Sun's motion \wrt\ the CMB \citep{HinWeiHil09}
allows a determination of the motion of the LG \wrt\ the CMB:
$622 \pm 35 \kms$  in the direction 
\lberr{272}{3}{28}{5}.
A deviation from the predicted value with that derived from the CMB
dipole would presumably arise from sources beyond the survey and would
thus have implications for large scale structure. As the tidal field
falls off as $r^{-3}$, to first order one can model additional
velocity contributions to the LG arising from sources outside the
survey volume as a dipole, with the magnitude of this dipole, or residual bulk flow, itself
being a test for cosmological models. Past studies constraining
cosmological parameters through comparison of predicted motion of the
LG using linear theory and that derived from the CMB include
\citet{ErdHucLah06}, and \citet{BilChoMam11}. Recently such an
analysis has been extended to the non-linear regime using a novel
orbit-reconstruction algorithm to predict motions of nearby objects
(\citealt{LavTulMoh10}).

In addition to reconstructing the motion of the Local Group, one can
explore the bulk motion of a large volume (typically the mean velocity
of a 50 \hmpc\ Gaussian-weighted window) as such motion probes the
amplitude of matter power spectrum on large scales. As previous
studies in this vein have found hints of excess power on large scales
({\it cf.} \citealt{WatFelHud09}), there has been interest in
performing such analyses using peculiar velocity surveys 
\citep[\eg][]{TurHudFel12, HonSprSta14}.

In this work we explore the methods used to self-consistently
reconstruct the real space density field from redshift space while
quantifying any biases intrinsic to this reconstruction method. We
then use the recently constructed large full-sky 2M++ catalogue
composed of 69,160 galaxy redshifts to measure $\beta$ to high
precision and constrain $f\sigma_{\rm{8}}$. We further explore the
growth of the LG's dipole as predicted by linear theory arising from
structures within 2M++, in addition to computing the bulk flow arising
from this survey.

This paper is organized as follows: in \S\ref{dfmethods} we briefly
review construction of the 2M++ catalogue, we discuss accounting for
incompleteness and functional dependence of galaxy bias on luminosity
when computing the density field, and we outline details of the
reconstruction procedure. In \S\ref{LGvel} we explore the growth
of the LG velocity amplitude arising from 2M++ as a function of survey
depth. In \S\ref{pvcompare} we discuss peculiar
velocity surveys used, the methods used in comparing predicted
velocities to measured velocities, as well as the results obtained
from such analyses. We discuss and compare our results to those from
recent literature in \S\ref{discussion}, and conclude in
\S\ref{conclusion}.
 
\section{Density Field Reconstruction}
\label{dfmethods}

Redshift surveys measure positions of objects in redshift-space. As
Equation (\ref{main}) requires real-space positions, we must first map
observed redshifts to real-space distances. The observed redshift,
$z\sbr{obs}$, is related to cosmological redshift resulting from the
expansion of the Universe, $z\sbr{cos}$, and that resulting from
peculiar velocities, $z\sbr{pec}$, through
\begin{equation}
(1+z\sbr{obs}) = (1+z\sbr{cos})(1+z\sbr{pec}),
\label{zrelations}
\end{equation}
\noindent where for non-relativistic peculiar motions $cz\sbr{pec}\simeq
v\sbr{pec}$.  Note that in the above if $z\sbr{obs}$ is 
corrected to some frame of reference, such as the CMB, then 
$v\sbr{pec}$ is the peculiar velocity with respect to that frame.
The comoving distance in the low-redshift Universe is
then related to the cosmological redshift through
\begin{equation}
H\sbr{0}R \simeq c\left(z\sbr{cos} - \frac{1+q\sbr{0}}{2}z\sbr{cos}^2
\right),
\label{z2dist}
\end{equation}
\noindent (\citealt{PPC}) where $R$ is the comoving distance,
$H_{\rm{0}}$ is the local value of the Hubble parameter, and where
$q\sbr{0}$ is the local value of the deceleration parameter, which is
given by $q = \Omega\sbr{m}/2 - \Omega_\Lambda$ for a flat
$\Lambda$CDM Universe. As real-space positions are dependent on
peculiar velocity predictions, which are themselves dependent on
real-space positions, mapping redshifts to real-space must be done
with care.

In this section, we discuss the 2M++ redshift catalogue, and outline
the procedure used in reconstructing comoving positions of the
galaxies therein. \S2.1 reviews construction of 2M++, and in \S2.2 we
outline how galaxy weights were computed to account for the fact that
2M++ is magnitude limited. In \S2.3 we briefly review the choice of
smoothing kernel. We discuss the procedure used to normalize the
smoothed density field to the same effective bias in \S2.4. In \S2.5
we outline the iterative scheme used to recover real-space positions
from redshift-space, and in \S2.6 we take a cosmographic tour through
the recovered density field.

\subsection{2M++ Redshift Compilation \label{2M++}}
The integral in Equation (\ref{main}) is over all space. In
reconstructing the velocity field of the local Universe, therefore,
clearly one would like a redshift survey that is very deep and as
close to all-sky as possible.  Two such catalogues which have been
used extensively in the past include the sparsely sampled IRAS Point
Source Catalogue Redshift Survey (PSCz, \citealt{SauSutMad00}), and
more recently, the shallower but more densely sampled Two-Micron
All-Sky Redshift Survey (2MRS, \citealt{HucMacMas11}).  In this work
we use a superset of 2MRS, dubbed 2M++, constructed by
\citet{Lavaux11}. This sample has greater depth than 2MRS, and
superior sampling than PSCz. The photometry is from
the Two-Micron-All-Sky-Survey (2MASS) Extended Source catalogue,
(2MASS-XSC, \citealt{SkrCutSti06}), an all-sky survey in the $J$, $H$
and $K_{\rm{S}}$ bands. Redshifts in the $K_{\rm{S}}$ band of the
2MASS Redshift Survey (2MRS) are supplemented by those from the Sloan
Digital Sky Survey Data Release Seven (SDSS-DR7,
\citealt{Abazajian09}), and the Six-Degree-Field Galaxy Redshift
Survey Data Release Three (6dFGRS, \citealt{JonReaSau09}). Data from
SDSS was matched to that of 2MASS-XSC using the NYU-VAGC catalogue
\citep{Blanton05}. As 2M++ draws from multiple surveys, galaxy
magnitudes from all sources were first recomputed by measuring the
apparent magnitude in the $K_{\rm{S}}$ band within a circular isophote
at 20 mag arcsec$^{-2}$. Following a prescription described in
\citet{Lavaux11}, magnitudes were then corrected for Galactic
extinction, cosmological surface brightness dimming and stellar
evolution. After corrections the sample was limited to
$K_{\rm{2M++}}\leq 11.5$ in regions not covered by 6dFGRS or SDSS, and
limited to $K_{\rm{2M++}}\leq 12.5$ elsewhere.

Other relevant corrections which were made to this catalogue include
accounting for incompleteness due to fibre-collisions in 6dF and SDSS,
as well as treatment of the \textit{zone of avoidance}
(ZoA). Incompleteness due to fibre-collisions was treated by cloning
redshifts of nearby galaxies within each survey region as described in
\citet{Lavaux11}.

\begin{figure*}
\begin{minipage}{0.65\linewidth}
  \raggedleft
  \includegraphics[width=\linewidth]{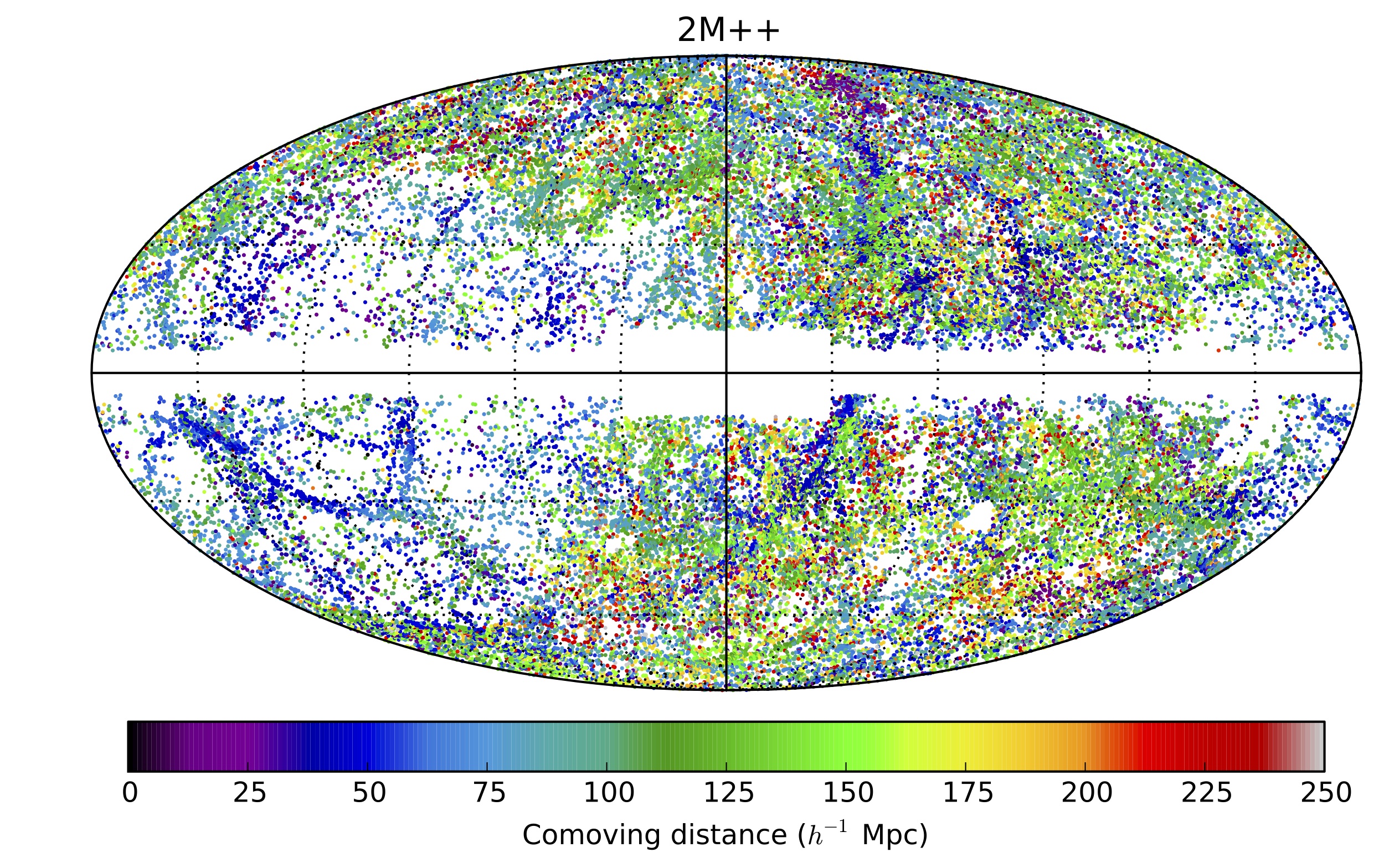}
\end{minipage}\begin{minipage}{0.33\linewidth}
  \raggedright
  \includegraphics[width=\linewidth]{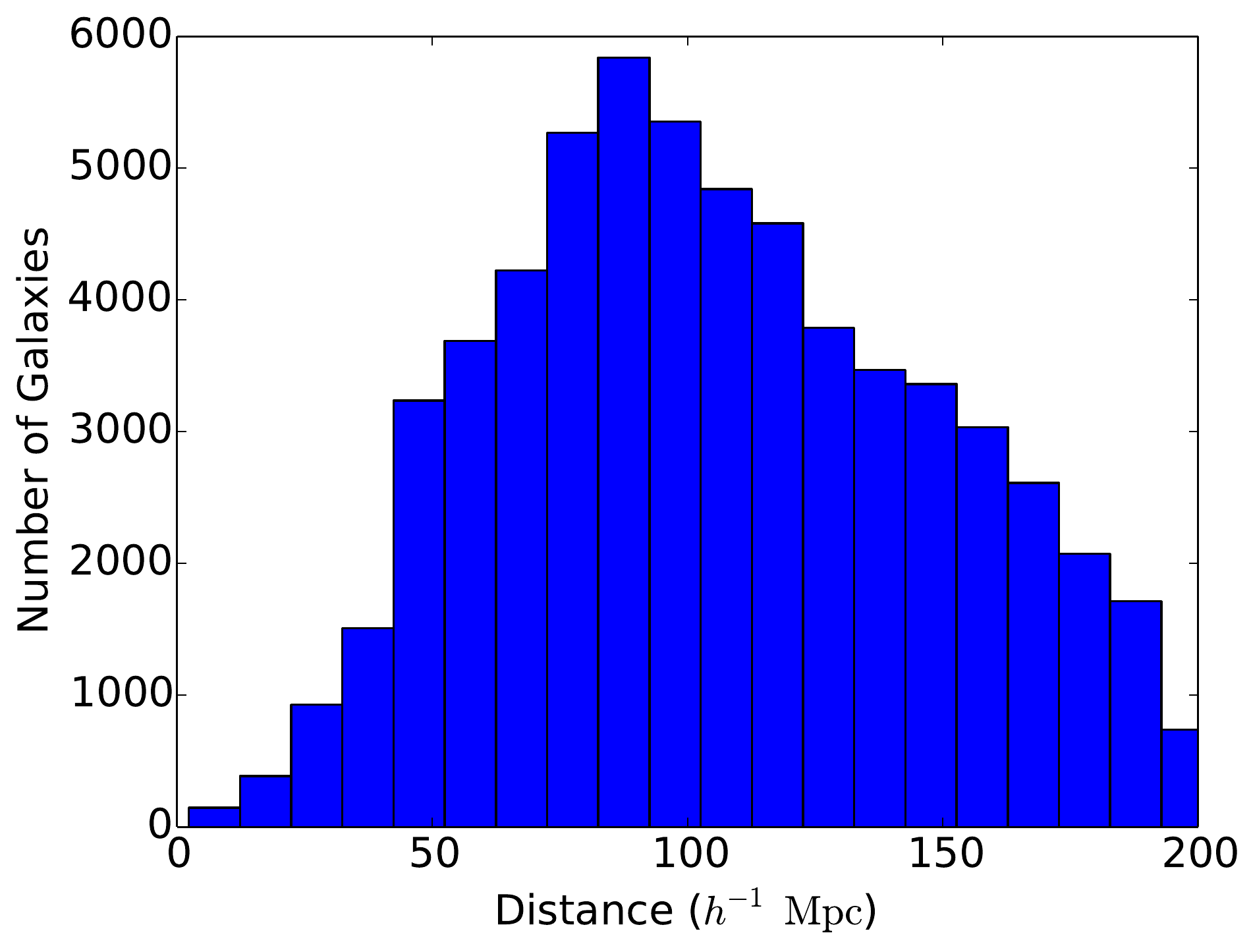}
\end{minipage}
\caption{Left: All galaxies in 2M++ with measured redshifts; blue galaxies are
nearest, red are farthest.  The 2MRS region ($K_{\rm{2M++}}\leq 11.5$) with lower density of galaxies is apparent. The Galactic centre is in the centre of the plot, and Galactic longitude increases to the left. Right: Histogram of galaxies in catalogue as a function of distance (bin width of 10\hmpc).}
\label{completeness}
\end{figure*}

In treating the ZoA, for Galactic longitudes in the range [30$^\circ$,
330$^\circ$], lower latitudes ($|b|<$5$^\circ$) were first masked and
then cloned with the redshifts from 2MRS in an equal-area strip 
just above the missing northern ($0^{\circ} < b < 5^{\circ}$) Galactic strip,
and, for the negative Galactic latitudes in the south, a strip below that was cloned.
Near the Galactic centre, for longitudes in the range [-30$^\circ$, 30$^\circ$], 
the wider Galactic latitude strip
$|b|<$10$^\circ$ was filled with the redshifts from 6dFGRS in a similar way. 
A histogram of distances is shown in Figure \ref{completeness}.

\subsection{Luminosity Function \& Galaxy Weights \label{LF}}
Before using the catalogue to construct the density field, we must
first account for survey incompleteness. In this section we provide a
summary of the method used to obtain the luminosity function fit to
the catalogue; this luminosity function is in turn used to in the
weighting scheme employed to account for incompleteness. For a
complete description of these calculations as applied to this
catalogue see \citet{Lavaux11}. The luminosity function used to
characterize the dataset is the Schechter function
\citeyearpar{Sch76}, which when written in terms of absolute
magnitudes is given by:
\begin{equation}
\begin{split}
&\Phi(M) = \\
&\;\;0.4\log (10)n^*10^{0.4(1+\alpha)(M^*-M)}\exp
\left(-10^{0.4(M^*-M)}\right),
\end{split}
\end{equation}
where $n^{*}$ is the density normalization, $M^{*}$ is the absolute
magnitude break, and $\alpha$ is additional power-law parameter to be
determined. Schechter function parameters are computed using likelihood
formalism, where the product of all conditional probabilities of observing
galaxies intrinsic magnitudes is maximized given their redshifts, Schechter
parameter values, and survey completeness at their specified angular positions
and distances.

Before computing weights we first discuss the different ways by which
we can model the galaxy density contrast. As we cannot compute the
mass density contrast of observed galaxies directly, we must use
either the number-density of galaxies, or their luminosity-density in
computing $\delta\sbr{g}$. In the context of linear biasing, our goal
is to create a galaxy density contrast field which most closely traces
the underlying total mass density contrast.  Although
luminosity-density may be a better proxy for stellar mass, and thus
for the underlying mass distribution of dark matter, we will consider
both schemes in this work.

To account for incompleteness, galaxies are weighted according to a
common prescription similar to that of \citet{DavHuc82}. In the case
of a number-density scheme, observed galaxies are weighted to account
for the number of galaxies not observed at a given distance due to the
magnitude limit of the survey. When using the galaxy number-density
for a single homogeneous redshift survey, galaxies are weighted by
\begin{equation}
w^N(r) = \frac{N\sbr{average}}{N_{\rm{observed}}(r)} = \frac{
\int_{L_{\rm{min}}}^{\infty}\Phi(L)\:dL } {\int_{4\pi r^2
f_{\rm{min}}}^{\infty}\Phi(L)\:dL}\:,
\end{equation}
when $4\pi d\sbr{L}^2 f_{\rm{min}} > L_{\rm{min}}$, and unity
otherwise. The flux limit, $f\sbr{min}$, corresponds to a
$K\sbr{s}$ band apparent magnitude limit of 11.5 for galaxies drawn
from 2MRS and 12.5 otherwise. The luminosity $L\sbr{min}$ used above
corresponds to a $K\sbr{s}$ band absolute magnitude of $-20$. Computed
weights used in this work additionally account for the inhomogeneous
incompleteness of 2M++ and a complete description of their
determination can be found in \citet{Lavaux11}.

\begin{figure}
\centering
\includegraphics[width=\linewidth]{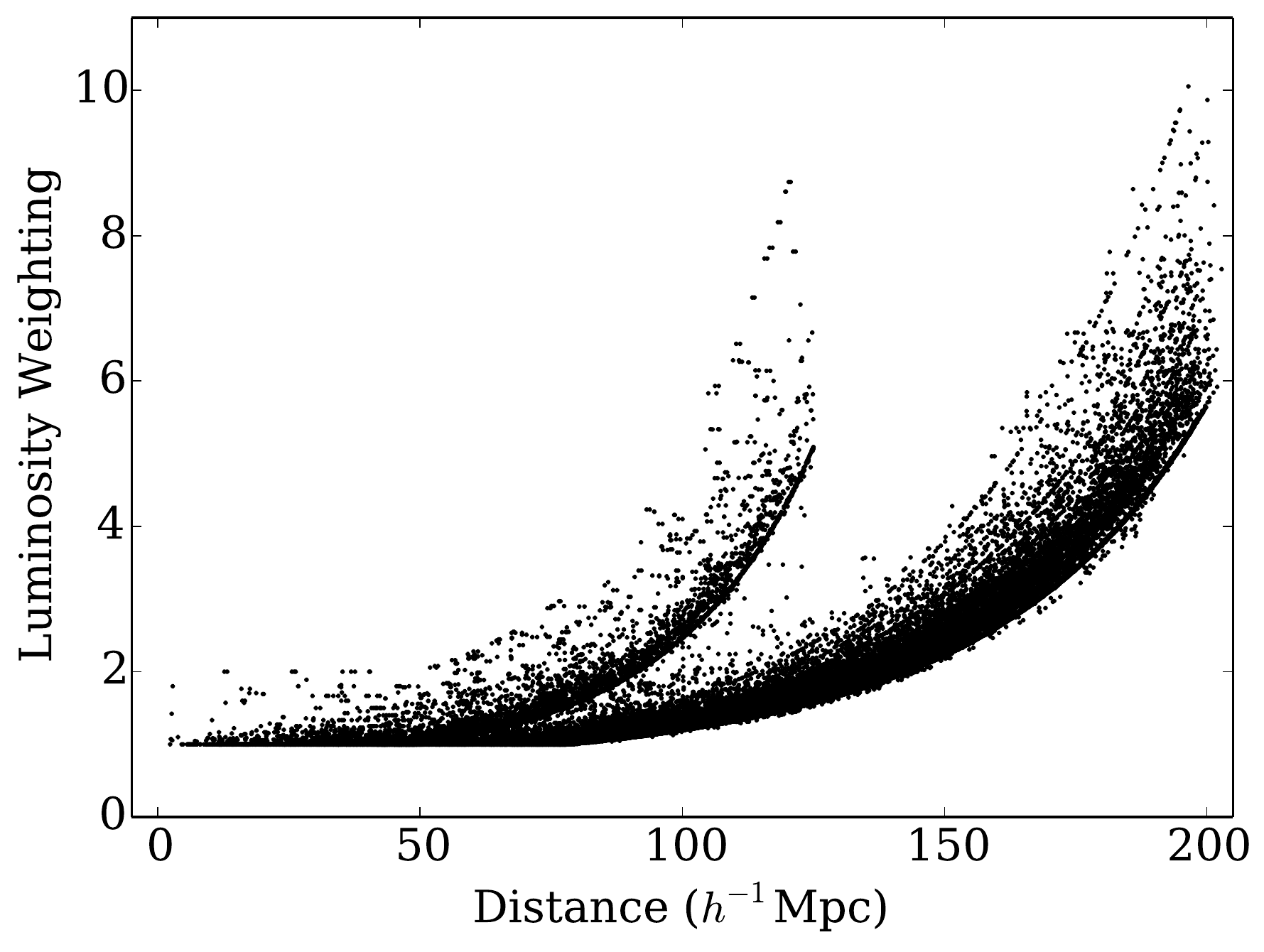}
\caption{Computed luminosity weights obtained from the procedure outlined in
\S\ref{LF}. As 2MRS is magnitude limited to $K\sbr{s} < 11.5$, weights rise more
sharply than for SDSS and 6dF galaxies which are magnitude limited to $K\sbr{s}
< 12.5$. The scatter in the weights at a given distance arise from varying
levels of redshift completeness across the sky.}
\label{plt:weights}
\end{figure}

In this paper, by default, we will use luminosity-density to compute
the galaxy density contrast. The weight assigned to each galaxy's
luminosity is again based on the fraction of the total luminosity
expected, given the magnitude limit of the survey, to the luminosity
one expects to observe at a given distance. Thus, for a single
homogeneous redshift survey, galaxy luminosities are weighted by
\begin{equation}
w^L(r) = \frac{L_{\rm{average}}}{L_{\rm{observed}}(r)} = \frac{
\int_{L_{\rm{min}}}^{\infty}L\:\Phi(L)\:dL } {\int_{4\pi r^2
f_{\rm{min}}}^{\infty}L\:\Phi(L)\:dL}\:.
\end{equation}
The calculated luminosity weights obtained for the best values of
cosmological parameters are shown in Figure \ref{plt:weights}.

\subsection{Smoothing \label{smooth}}
In order to use Equation (\ref{main}), the density field must first be
sufficiently smooth for linear theory to apply. The optimal scale on
which to smooth the data was determined by comparing velocities from
an N-body simulation to predictions obtained through linear theory
using different smoothing lengths.  Smoothing the density contrast
with a 4 \hmpc\ Gaussian was found to be the best compromise in
minimizing the scatter in predicted velocities vs. simulation
velocities, while simultaneously returning an unbiased slope in the
comparison of observed nonlinear velocities (from the simulation) with
the linear theory predictions from a smoothed, reconstructed density
field of halos.  These comparisons are discussed in greater detail in
Appendix A.

\subsection{Accounting for Magnitude Dependence of Galaxy-Matter
  Bias}\label{bias}
For magnitude limited surveys such as 2M++, the mean luminosity of
observed galaxies increases with depth. As galaxy-matter bias has been
found to increase with luminosity, this means that objects observed at
higher redshift are on average more biased than those observed
nearby. In this section we account for this effect by rescaling the
density field to the same effective bias. We do so using the bias
model of \citet{Westover07} in which bias is a function of luminosity.
By comparing the correlation function of 2MASS volume-limited
subsamples, converting the binned absolute magnitude to a luminosity
and defining $b/b^* = (\xi(s)/\xi_{\rm{fid}}(s))^{1/2}$,
\citet{Westover07} found $b/b^* = (0.73 \pm 0.07) + (0.24 \pm
0.04)L/L^*$, where $b^*$ is the bias of an $L^*$ galaxy. This result
is consistent with that of \citet{Norberg01} and \citet{Tegmark02},
who did similar analyses using projected correlation functions of
2dFGRS and the SDSS power spectrum, respectively.

Since bias is a function of luminosity, this means that a naive
computation of the density field using a magnitude limited survey
would lead to a larger effective bias at larger distances. As our end
goal is to compare predicted velocities with measured velocities and
determine $f/b$, we must first correct the density contrast field by
normalizing the field to the same effective bias. The effective
number-weighted bias is computed as follows:
\begin{equation}
b_{\rm{eff}}^{N}(r) = \frac{\int_{4\pi r^2f_{\rm{min}}}^{\infty}
b(L)\:\Phi(L)\:dL}{\int_{4\pi r^2f_{\rm{min}}}^{\infty} \Phi(L)\:dL} =
\psi^N(r)b^*.
\end{equation}
Using a luminosity-weighting scheme in computing the density contrast,
the effective luminosity-weighted bias is given by:
\begin{equation}
b_{\rm{eff}}^{L}(r) = \frac{\int_{4\pi r^2f_{\rm{min}}}^{\infty}
b(L)\:L\:\Phi(L)\:dL}{\int_{4\pi r^2f_{\rm{min}}}^{\infty} L\:\Phi(L)\:dL} =
\psi^L(r)b^*.
\end{equation}

\begin{figure}
\centering
\includegraphics[width=\linewidth]{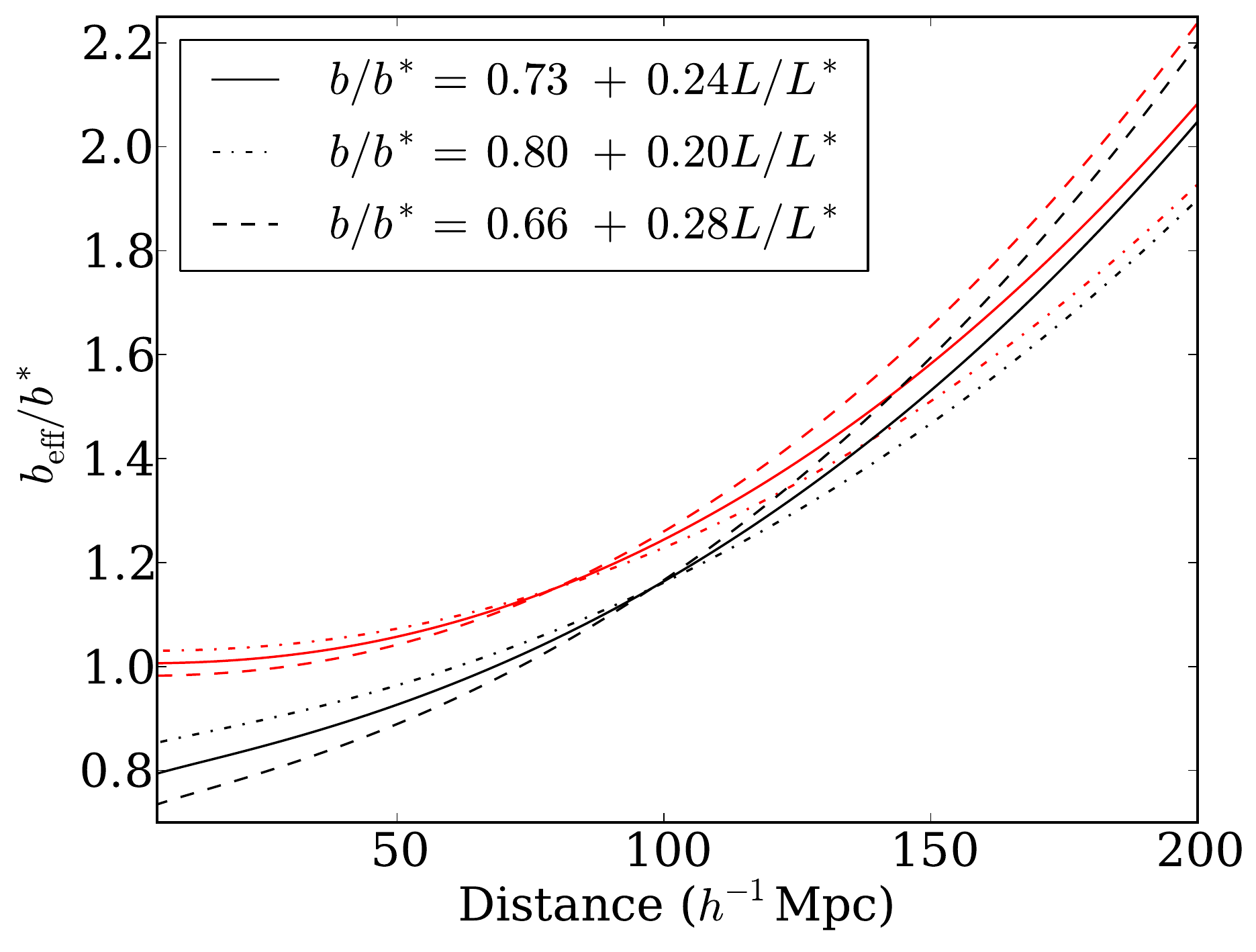}
\caption{Number-weighted (lower) and luminosity-weighted (upper)
  effective bias as a function of distance for the most significant
  1$\sigma$ deviation of parameters from the scaling relation $b/b^* =
  (0.73 \pm 0.07) + (0.24 \pm 0.04)L/L^*$. Plots are obtained using
  the parameter values $\alpha = -0.85$ and $M^* = -23.25$ for the
  Schechter luminosity function.}
\label{bias_plt}
\end{figure}

Using the functional form of $b(L)$ quoted above from Westover (2007),
this normalization procedure was applied to the 2M++ density contrast
fields. Both the number-weighted and luminosity-weighted effective
bias for $\alpha = -0.85$, $M^* = -23.25$, and a magnitude limit of
12.5 are shown in Figure \ref{bias_plt}. We can then rewrite Equation
(\ref{main}) as
\begin{equation}
\mathbf{v}(\mathbf{r}) = \frac{\beta^*}{4\pi}\int
d^3\mathbf{r}'\delta_{\rm{g}}^*(\mathbf{r}')\frac{(\mathbf{r}'
- \mathbf{r})}{|\mathbf{r}' - \mathbf{r}|^3}\ ,
\label{bstarlineartheory}
\end{equation}
where we have defined $\beta^* \equiv f(\Omega)/b^*$, and
$\delta_{\rm{g}}^*(r) \equiv b^*\delta(r) =
\delta_{\rm{g}}(r)/\psi(r)$.

\subsection{Reconstruction Procedure \label{procedure}}
The weighted galaxies from 2M++ have measured redshifts $cz$, and not
precise distances $r$.  However, application of \eqref{main} requires
distances as opposed to redshifts.  We refer to the inverse problem of
determining the positions from redshifts as ``reconstruction.''
Reconstruction was accomplished via an iterative procedure modeled on
that of \citet{Yahil91}. Objects were first grouped using the
``Friends-of-friends" algorithm \citep{HucGel82}, and then placed at
the mean of their group redshift distance to suppress the
``Fingers-of-God'' effect. Gravity was then ``adiabatically'' turned
on by increasing \b\ $\equiv f(\Omega\sbr{m})/b^*$ from 0 to 1 in
steps of 0.01. The reconstruction took place in the LG frame, and on
each iteration the following steps were taken:
\begin{enumerate}
\item A Schechter luminosity function is fitted to the data using the
  likelihood formalism discussed in \S\ref{LF}. The LF is in turn used
  to compute either the luminosity or number weights following the
  procedure discussed in \S\ref{LF}.
\item Galaxies from 2MRS with distances greater than 125 \hmpc\ are
  assigned a weight of zero. Galaxy properties, including newly
  computed weights are then cloned to account for incompleteness and
  fill the ZoA as described in \S\ref{2M++}.
\item Number weighted galaxies or their weighted luminosities within
  200 \hmpc\ are then placed on a grid.\item The density constrast field is then computed and normalized to
  the same bias, $b^*$, as described in \S \ref{bias}. The field is in
  turn smoothed with a Gaussian kernel of width 4 \hmpc.
\item Using Equation (\ref{bstarlineartheory}) the density contrast
  field is then used to obtain predicted peculiar velocities for all
  objects in the catalogue.
\item In conjunction with measured redshifts, predicted peculiar
  velocities projected on to the line-of-sight are then used to
  predict comoving distances using Equations (\ref{zrelations}) and
  (\ref{z2dist}).
\item The previous five predictions for a galaxy's distance are then
  averaged to suppress oscillations arising from triple valued
  regions, \ie regions of high-density near which there are multiple
  solutions for distance given redshift (discussed further in Appendix
  B). The averaged distance is in turn assigned to the galaxy
  and used to recompute the galaxy's absolute magnitude. Computed
  distances and magnitudes are then used in the subsequent iteration.
\end{enumerate}
Catalogues containing updated distances, luminosities and weights were
saved at each iteration (corresponding to increasing values of
$\beta$) in addition to the computed density and velocity fields.

It should be noted that this iterative reconstruction procedure was
found to be unbiased in determining the best fit value of \b\ when the
full analysis was run on an N-body simulation using a $\Lambda$CDM
cosmology (see Appendix A for more details).

\subsection{Cosmography}
\begin{figure*}
\centering
\includegraphics[width=\linewidth]{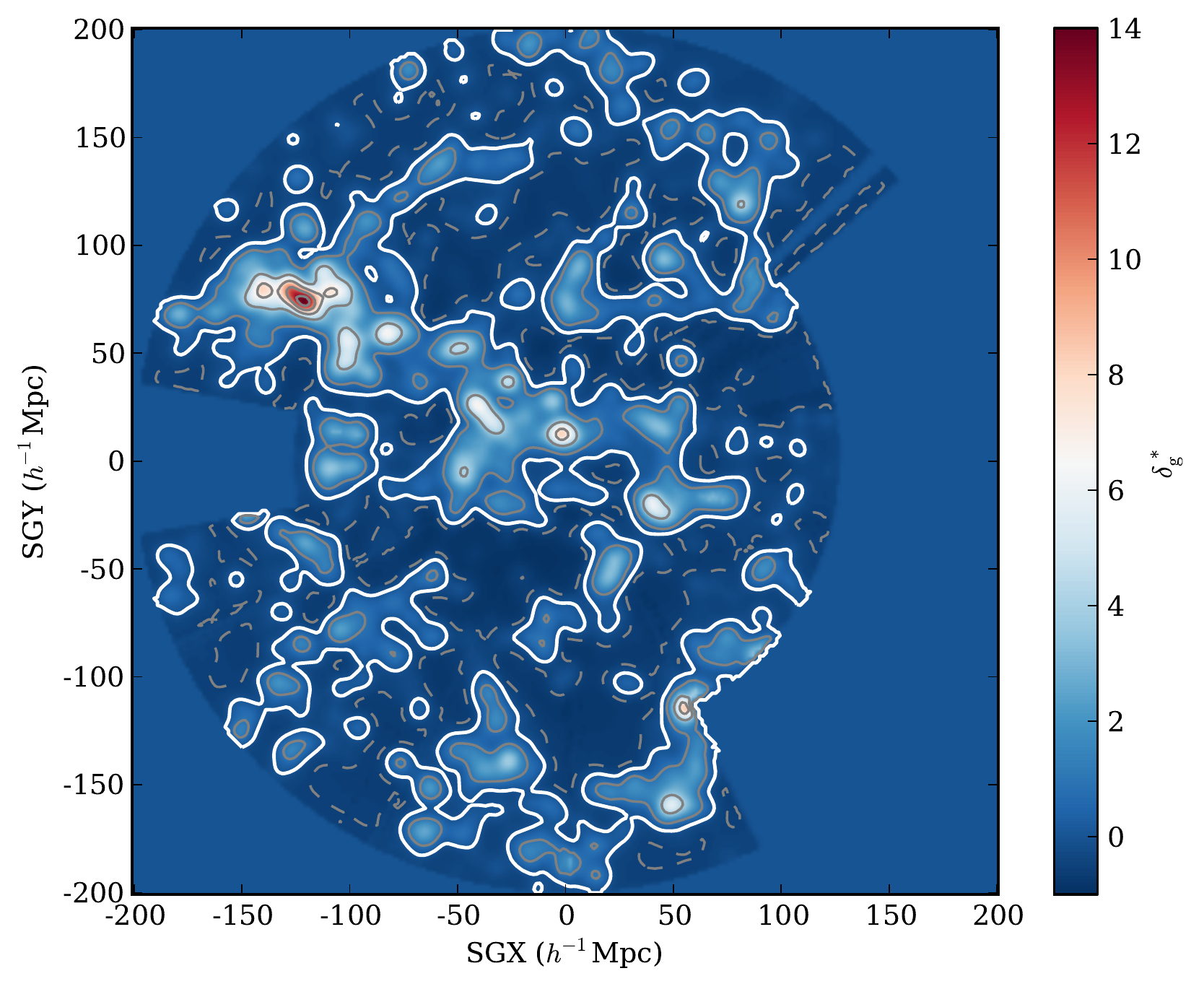}
\caption{The Supergalactic Plane (SGZ$=0$) of the 2M++ luminosity-weighted galaxy density contrast field, reconstructed with $\beta^* =
  0.43$ smoothed with a Gaussian kernel of radius 4 \hmpc. The dashed
  contour is $\delta\sbr{g}^* = -0.5$, the bold white contour is
  $\delta\sbr{g}^* = 0$, and successive contours thereafter increase
  from 1 upwards in steps of 3. The Galactic plane runs roughly along the SGY$=0$ axis. The Shapley
Concentration is located at (SGX, SGY) $\simeq$ (-125, 75) \hmpc,  the Virgo Supercluster
directly above the LG, the Hydra-Centaurus Supercluster at (-40, 20)
\hmpc, and the Perseus-Pisces Supercluster is at (40, -30)
\hmpc. The density field is shallower at positive SGX because this region is only covered by the 2MRS, 
whereas the rest of the plane is covered by the deeper 6dFGRS and SDSS.}
\label{SGP}
\end{figure*}

\begin{figure*}
\centering
\begin{minipage}{.49\linewidth}
  \centering
  \includegraphics[width=\linewidth]{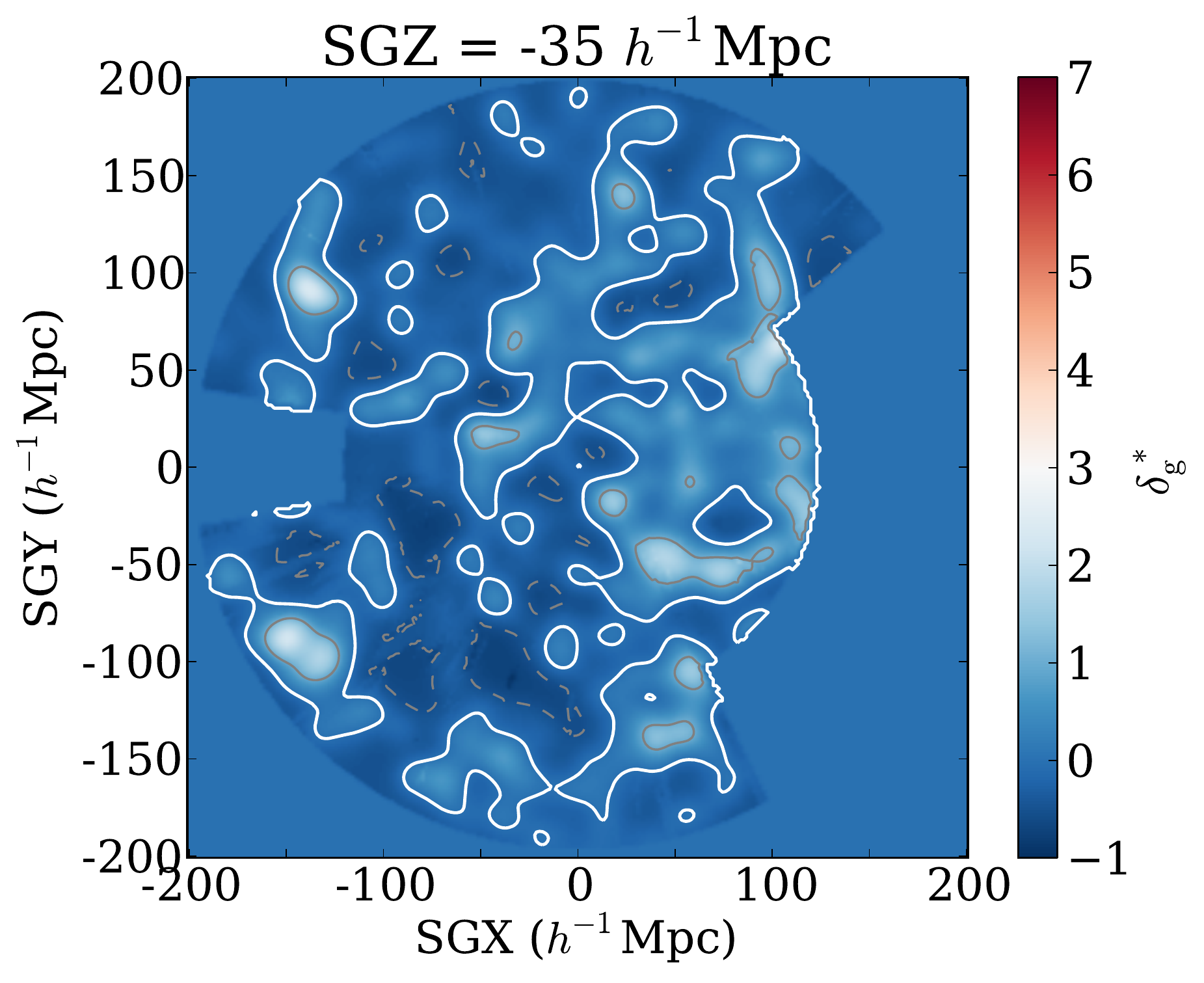}
\end{minipage}\begin{minipage}{.49\linewidth}
  \centering
  \includegraphics[width=\linewidth]{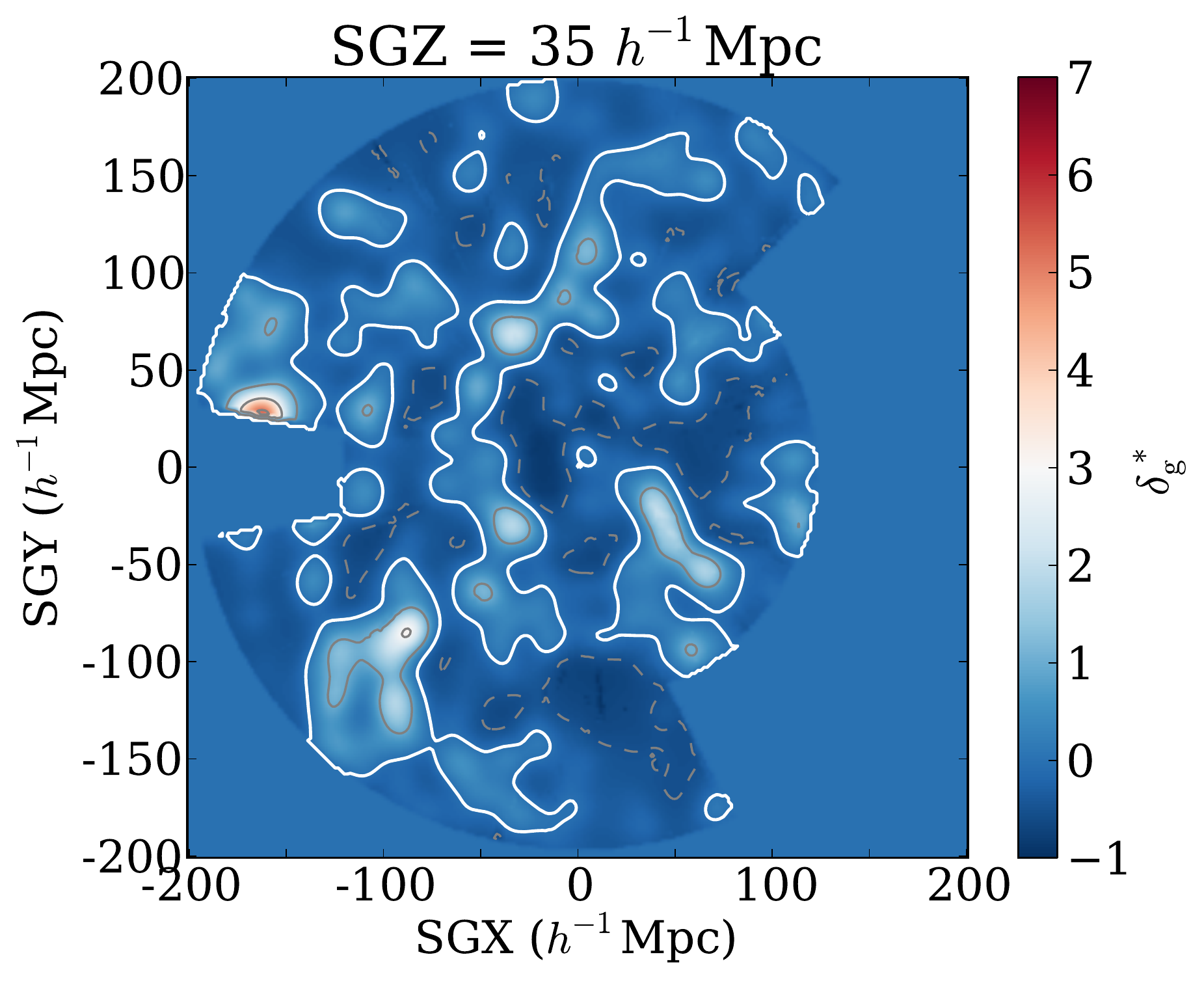}
\end{minipage}
\begin{minipage}{.49\linewidth}
  \centering
  \includegraphics[width=\linewidth]{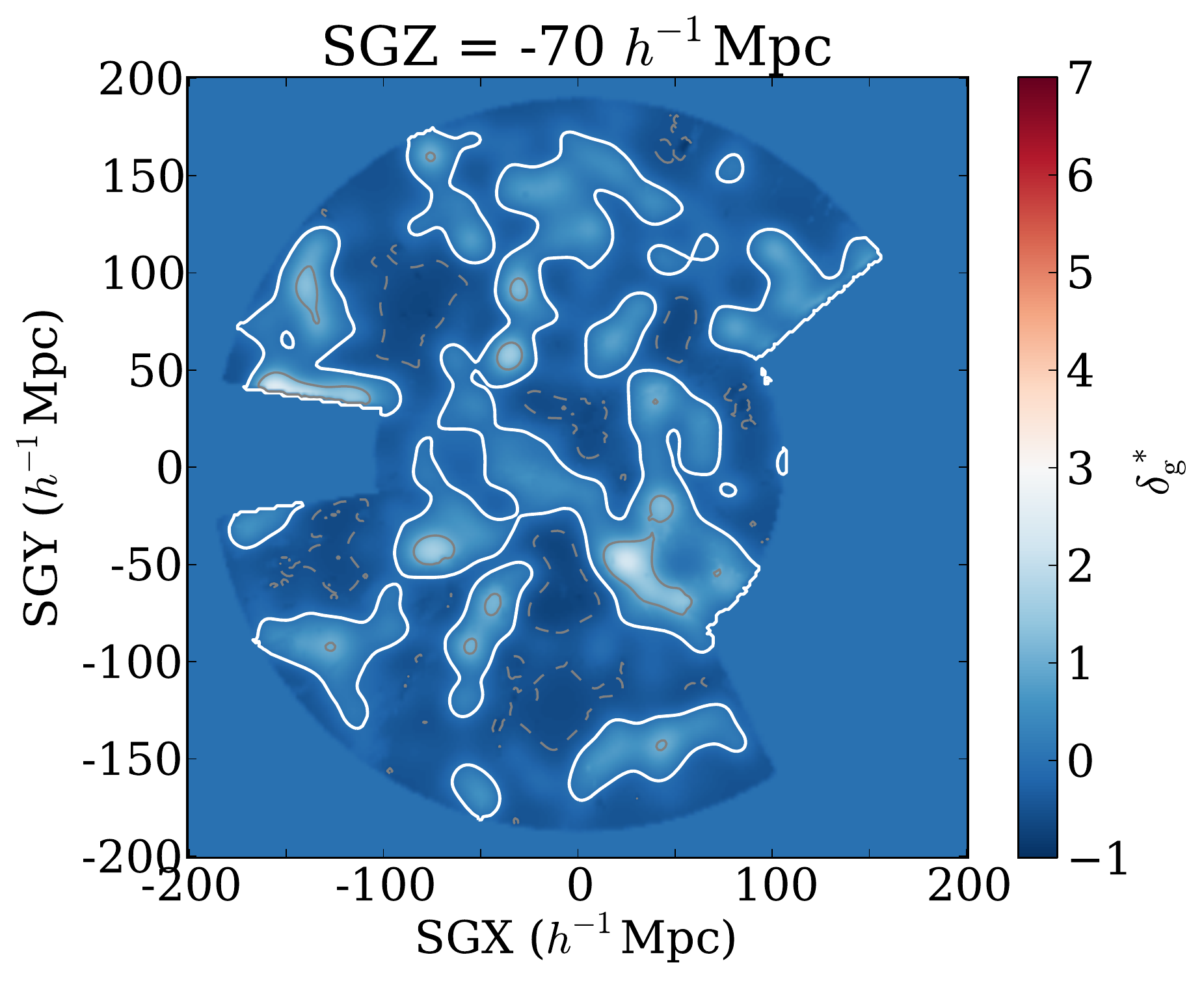}
\end{minipage}
\begin{minipage}{.49\linewidth}
  \centering
  \includegraphics[width=\linewidth]{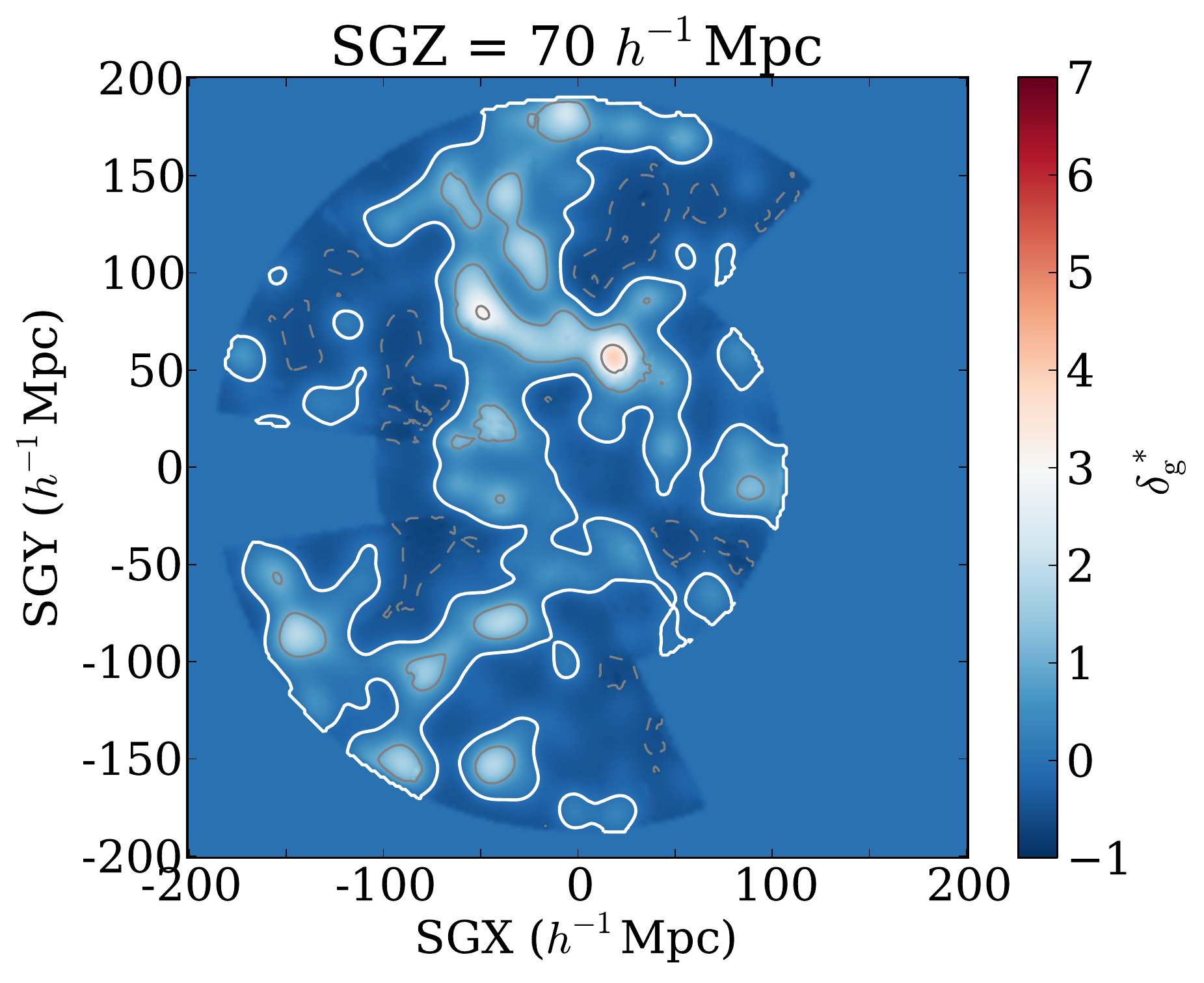}
\end{minipage}
\begin{minipage}{.49\linewidth}
  \centering
  \includegraphics[width=\linewidth]{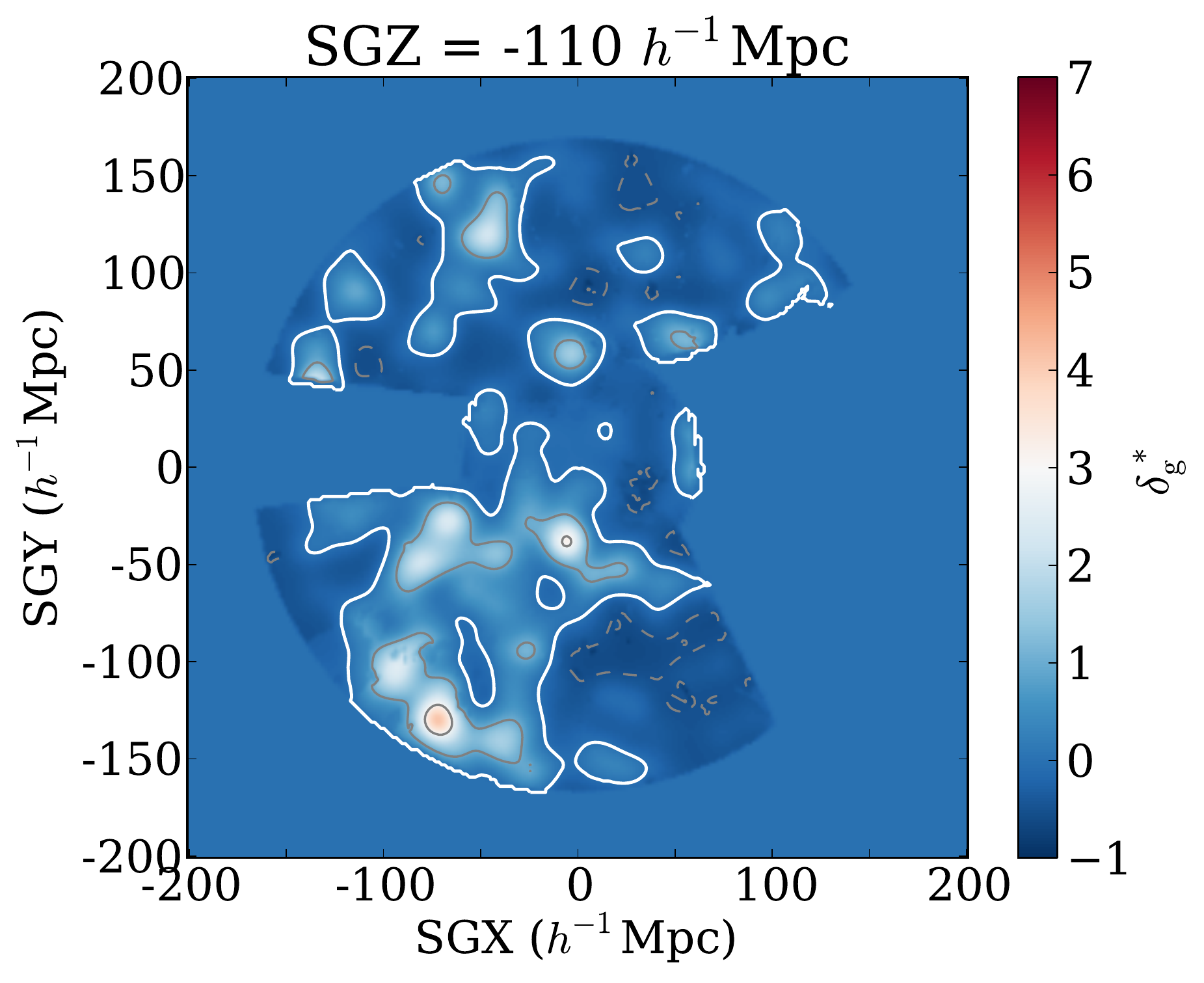}
\end{minipage}
\begin{minipage}{.49\linewidth}
  \centering
  \includegraphics[width=\linewidth]{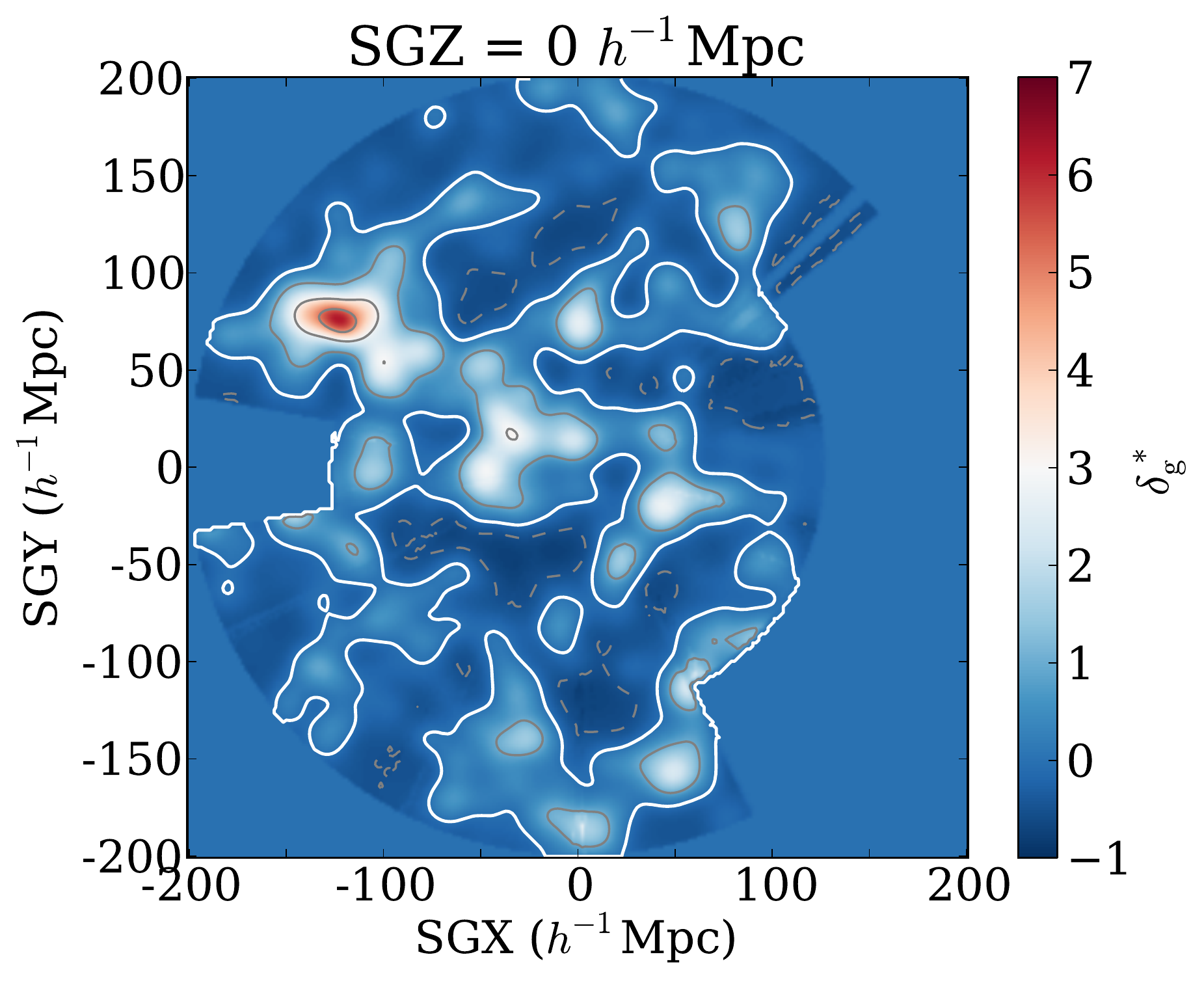}
\end{minipage}
\caption{The 2M++ density field at various slices of SGZ. The density
  field was smoothed with a 7 \hmpc\ Gaussian kernel. 
  The SGZ $=-110 \hmpc$ panel shows the Horologium-Reticulum supercluster.
  The dashed contour is $\delta\sbr{g}^* = -0.5$, the bold white contour is
  $\delta\sbr{g}^* = 0$, and successive contours thereafter increase
  from 1 upwards in steps of 2.}
\label{SGplots}
\end{figure*}

Figure \ref{SGP} shows the luminosity-weighted density field of the
Supergalactic Plane for $\beta^* = 0.43$, smoothed with a 4 \hmpc\
Gaussian kernel. The incomplete coverage due to the lower magnitude
limit of 2MRS is clearly visible in this figure beyond SGX $\simeq$
125 \hmpc. The most prominent overdensity in this plane is the Shapley
Concentration located at (SGX, SGY) $\simeq$ (-125, 75) \hmpc. Other
notable structures in Figure \ref{SGP} include the Virgo Supercluster
directly above the LG, the Hydra-Centaurus Supercluster at (-40, 20)
\hmpc, and the Perseus-Pisces Supercluster (40, -30)
\hmpc. Additional slices through SGZ are shown in Figure
\ref{SGplots}, though smoothed on a 7 \hmpc\ scale to enhance the
contrast of large overdensities. For instance, Horologium-Reticulum
Supercluster can be readily observed at SGZ $\simeq$ -112 \hmpc, SGX
$\simeq$ -70 \hmpc, SGY $\simeq$ -140 \hmpc. The supergalactic plane
is also shown with this smoothing for comparison.

\section{2M++ Predicted Peculiar Velocity of the Local Group}\label{LGvel}
The velocity of the LG as predicted by linear theory for an ideal
distance-limited catalogue is given by:
\begin{equation}
\mathbf{v}_{\rm{LG}} = \frac{\beta}{4\pi}\int_{\rm{0}}^{R_{\rm{max}}}
d^3\mathbf{r}'\delta_{\rm{g}}(\mathbf{r}')\frac{\mathbf{r}'}{r'^3} +
\mathbf{V_{\rm{ext}}}\ ,
\label{LG1}
\end{equation}
where $\mathbf{V_{\rm{ext}}}$ encapsulates contributions from beyond
$R_{\rm{max}}$, and to first order, can be approximated as a
dipole, or ``residual'' bulk flow. For a realistic flux-limited catalogue, we do not detect a
continuous distribution of matter but a finite number of galaxies. As
a result, our estimate of the velocity of the LG using 2M++ is subject
to shot noise. To estimate the effect of shot noise on our predicted
motion of the LG, we computed the standard deviation in each of the
components of 500 bootstrap samples. The shot noise in the amplitude
of the LG's motion from this analysis was found to be 56 \kms.

Under the assumption that the observed CMB dipole arises from the
motion of the LG, there has been much debate as to the structures
sourcing this motion. Most recent work has made use of 2MASS-XSC or
2MRS in reconstructing the motion of the LG, and there has yet to be
consensus on the distance at which the LG's motion coincides with that
derived from the CMB. \citet{ErdLahHuc06} argue that more than 70\% of
the LG's motion results from structures within 50 \hmpc, such as
Hydra-Centaurus Supercluster. While others argue for convergence at
distances greater than $\sim$120 \hmpc\ such as \citet{LavTulMoh10}
using Monge-Amp\`ere-Kantorovich orbit-reconstruction method, or
\citet{BilChoMam11}, who explored the convergence of the 2MASS dipole
moment of the angular distribution of galaxies as a function of the
limiting flux of the sample. As 2M++ is a superset of 2MRS and
contains redshift measurements up to a magnitude limit of $K_{\rm{s}}
= 12.5$, this data is well suited to examine the influence of
structures beyond 120 \hmpc\ on the velocity of the Local Group.

\begin{figure*}
\centering
\begin{minipage}{.5\linewidth}
  \centering
  \includegraphics[width=\linewidth]{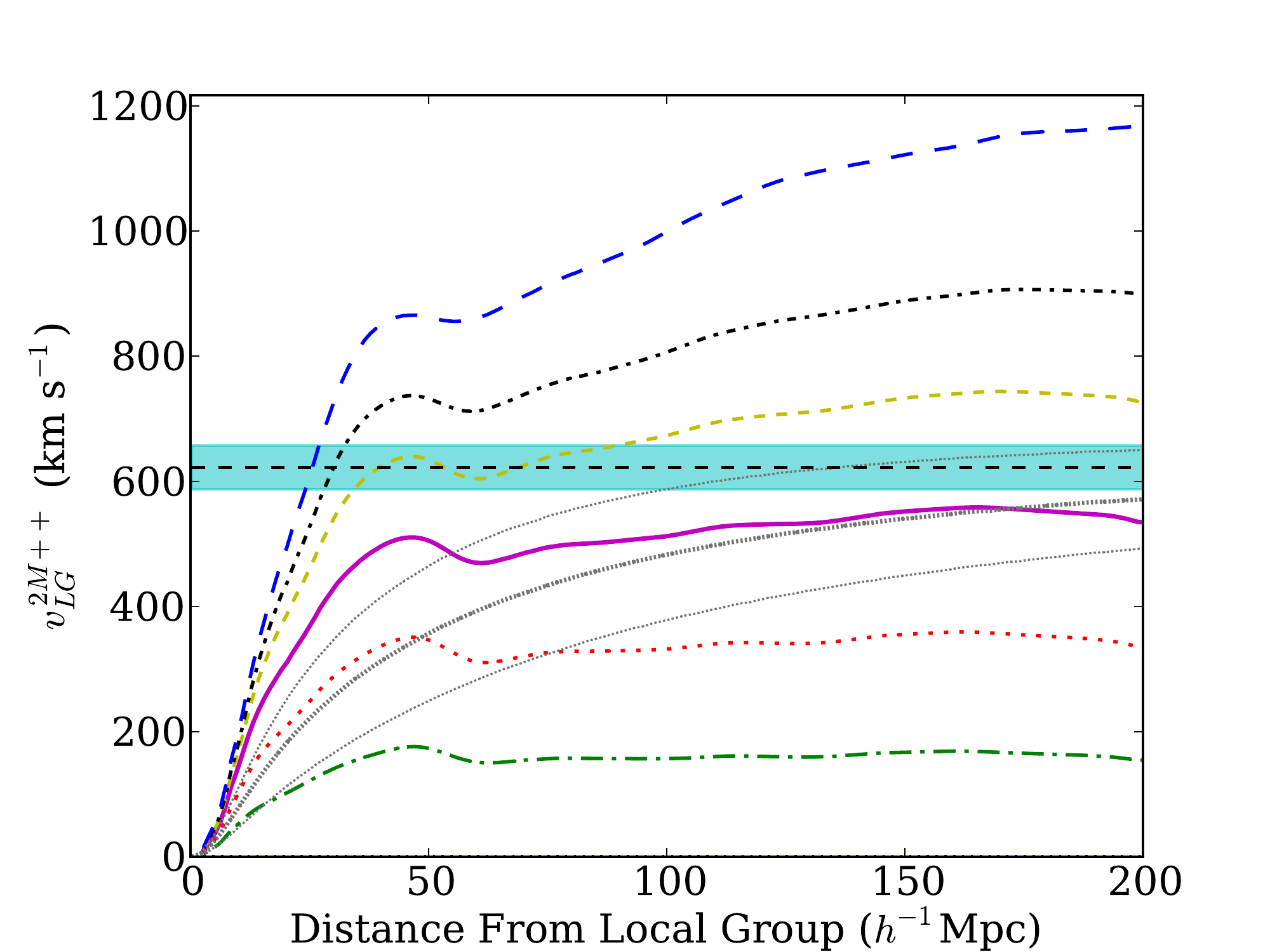}
\end{minipage}\begin{minipage}{.5\linewidth}
  \centering
  \includegraphics[width=\linewidth]{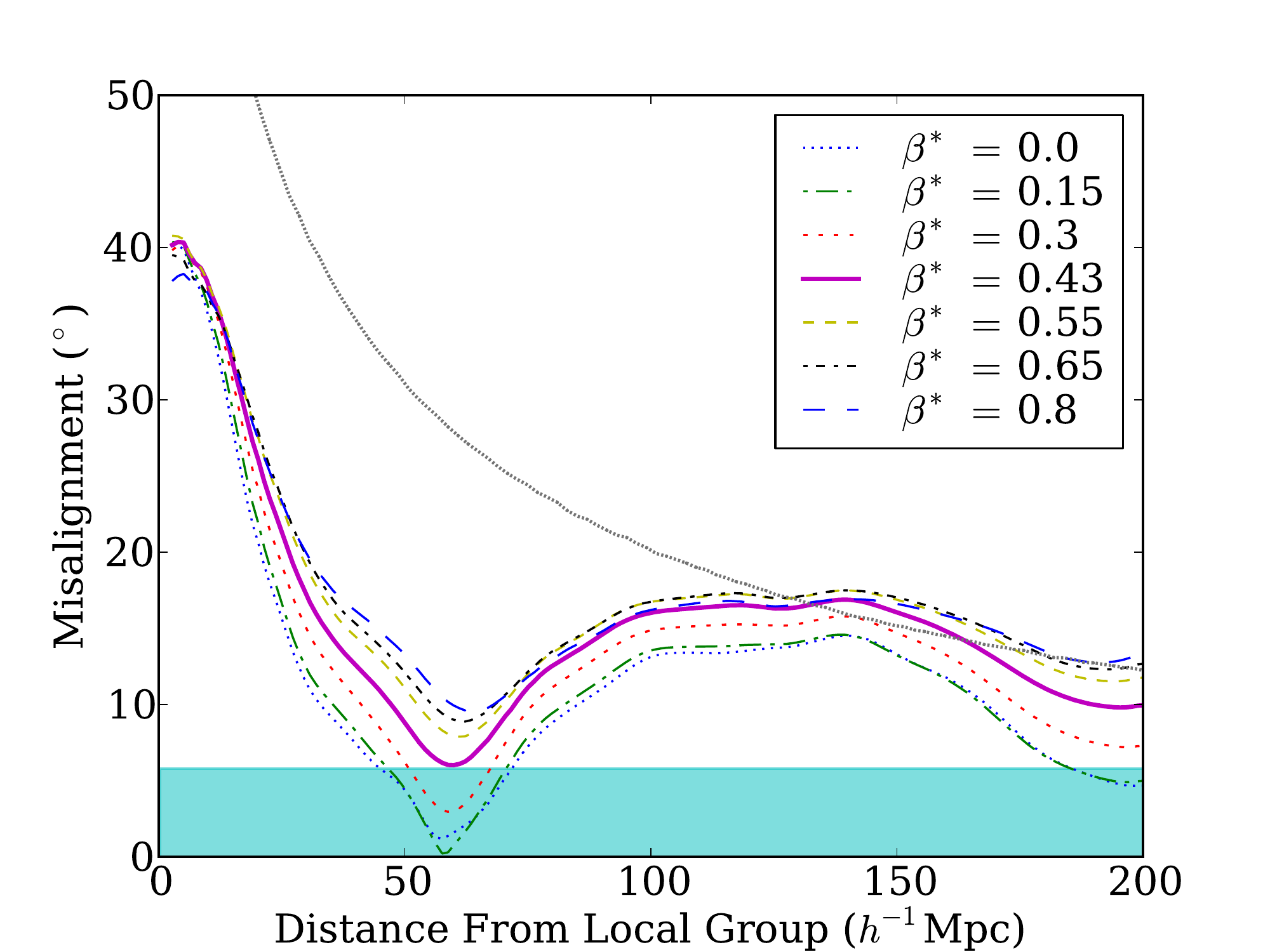}
\end{minipage}
\caption{Left: Growth of LG velocity amplitude as predicted by linear
  theory for successively larger concentric spheres. The non-monotonic
  curves correspond to the 2M++ predicted growth of the LG amplitude
  for different values of $\beta^*$ as indicated in the legend. The
  solid purple curve is for our best fit value $\beta^* = 0.43$. The
  smooth thick grey line corresponds to the expected velocity
  amplitude for a survey of a depth indicated by the x-axis in a
  $\Lambda$CDM WMAP9 cosmology.  The thin grey lines indicate the 68\%
  uncertainties due to cosmic variance.  The shaded cyan band
  corresponds to the velocity of the Local Group and its 68\%
  uncertainties as inferred from the CMB dipole. Right: The misalignment of the LG predicted direction of the velocity
  arising from 2M++ with that derived from the CMB ($l = 272^\circ$,
  $b = 28^\circ$). The smooth grey curve indicates the 68\%
  uncertainty on the alignment expected in the WMAP9
  cosmology.} \label{LG}
\end{figure*}

Using density fields for different values of \b\ that were obtained
throughout the iteration procedure discussed in \S \ref{procedure},
the velocity field of increasingly larger concentric spheres centered
on the LG was computed. The direction and amplitude of the LG velocity
as a function of distance for different values of \b\ was then
obtained. The amplitude of the LG velocity as predicted by 2M++ for
different values of \b\ is shown in Figure \ref{LG}a. The expected
agreement between predictions using linear theory for a survey of a
certain depth with values derived from the CMB are plotted for
comparison. This conditional probability assumes a $\Lambda$CDM WMAP9
cosmology. A derivation of this conditional velocity can be found in
Appendix A of \citet{LavTulMoh10}, and is based on
\citet{LahKaiHof90}. The convergence of the direction of LG velocity
with that derived from the CMB ($l = 272^\circ \pm 3^\circ$, $b =
28^\circ \pm 5^\circ$) is plotted in Figure \ref{LG}b. For our best
fit value of \b\ $=$ $0.43$ from \S\ref{pvcompare} below, the
misalignment is $10^\circ$. This misalignment angle is significantly
better than those found by past studies using the shallower 2MRS, such
as the $21^\circ$ misalignment found by \citet{ErdHucLah06},
$19^\circ$ found by \citet{BilChoJar11}, or the $\sim 45^{\circ}$
misalignment found by \citet{LavTulMoh10}.

\section{Peculiar Velocity Comparisons}

In principle it is possible to constrain cosmological parameters by comparing 
the growth of predicted motion of the LG with its observed motion.
This is difficult in practice because one expects contributions from 
sources beyond the survey limit, 
which we have modeled here as a residual dipole \Vext.
There is therefore a degeneracy between \b and
\Vext\ parameters appearing in equation \eqref{LG1}.
Very early studies of the growth of the LG gravity dipole observed the flatness at large radii 
from data similar to that shown in Figure \ref{LG}, 
therefore assumed that \Vext\ was negligible, and hence solved for \b. 
This procedure, however,  leads to a value of $\beta$ which is biased high, 
since \Vext\ and the integral in \eqref{LG1} are correlated. 
An alternative approach is to constrain \Vext\ by assuming a
cosmological model of density fluctuations, but this then makes the
exercise model-dependent. 

A better approach is to break the degeneracy by measuring
peculiar velocities of galaxies or groups other than the LG. 
In this section, we discuss comparisons between the predicted and observed
motions of two samples of galaxies, derived from the Tully-Fisher (TF)
relation and from Type Ia supernovae. 

\label{pvcompare}
\subsection{Peculiar Velocity Surveys}
\subsubsection{SFI++}
SFI++ (\citealt{SprMasHay08}) builds primarily on Spiral Cluster
I-band (SCI) and Spiral Field I-band (SFI) samples and uses a mixture
of 21-cm line profile widths and optical rotation curves in
determining the I-band Tully-Fisher (TF) relation from a subset of 807
galaxies in the fields of 31 clusters and groups
\citet{MasSprHay06}. From the derived TF relation they in turn
determine the peculiar velocities of 5780 galaxies. Upon removing
galaxies without high-quality width measurements and those that are
located beyond the volume covered by 2M++, SFI++ can be divided in to
two subsets of 2583 field galaxies and 735 galaxy groups.

As noted by \citet{DavNusMas11}, the SFI++ TF relation has a kink in
the faint end ($M > -20$), and an asymmetric distribution of outliers
about the expected velocity width parameter $\eta\equiv\log(W)-2.5$
(\citealt{SprMasHay08}). As we will be fitting for the inverse TF
relation, we will account for the outliers and deviation from
linearity of the relation by excluding galaxies with redshift-distance
magnitudes fainter than -20. We then iteratively compute the TF
relation parameters and remove those with a velocity width that
deviates by more than 0.2 in $\eta$ (3.8$\sigma$) from the relation,
until derived parameter values converge. Selection on both magnitude
and velocity width resulted in the rejection of 503 field galaxies and
137 galaxy groups.  Furthermore, when comparing predicted velocities
from 2M++ with those from SFI++, the remaining objects which were
found to differ by more than 3.5$\sigma$ with all velocity fields
obtained through the reconstruction procedure were rejected
(0.6\%). The final sample was composed of 2067 field galaxies and 595
galaxy groups. The typical or characteristic depth of the sample can
be quantified by a weighted mean distance, where the weights are the
inverse square of the uncertainties. This yields depths of of 42
\hmpc\ and 25 \hmpc\ for field and group samples respectively.

For the TF relation, we will perform the fit two different ways. The first is a direct maximum-likelihood fit to the observed linewidths: the VELMOD method of \citet{WilStrDek97}, as described in \secref{velmod} below.  The second method uses the estimated distances as given by \citet{SprMasHay08}, but corrected for the fact that their peculiar velocities were obtained under the assumption that $cz\sbr{obs}=H_{0}R + v\sbr{pec}$. Specifically,  we use the analytic relation from Equation (\ref{zrelations}) to obtain velocities from measured positions.

\subsubsection{First Amendment Supernovae}
The First Amendment (A1) catalogue Type Ia Supernovae (SNe) datasets
compiled by \cite{TurHudFel12}. A1 is composed of SNe within 200
\hmpc\ and draws 34 SNe from \citet{Jha07}, 185 from \citet{Hicken09}
and 26 from \citet{Folatelli10}. Of these 245 SNe, 237 are within the
volume spanned by 2M++, and have an uncertainty-weighted depth of 31
\hmpc.

\subsection{Velocity-Velocity Comparisons \label{vv}}
For the SFI++ subsets we use the distances as determined in
\citet{SprMasHay08} which have not been corrected for Malmquist
bias. We similarly do not use Malmquist bias corrected A1
distances. Of the comparison methods discussed below, VELMOD is
unaffected by inhomogeneous Malmquist bias, and accounts for
homogeneous Malmquist bias in the likelihoods. The Forward Likelihood
method discussed accounts for both homogeneous and inhomogeneous
Malmquist bias in the likelihoods. The simple $\chi^2$ comparison,
however, neither accounts for homogeneous nor inhomogeneous Malmquist
bias, and as such, the results yielded from this analysis are taken to
be biased.

\subsubsection{VELMOD}
\label{sec:velmod}
VELMOD is a rigorous maximum likelihood method first proposed and
implemented by \citet{WilStrDek97} and described further in
\citet{WilStr98}. It is a velocity-velocity comparison method used to
fit for the TF relation parameters (zero-point, slope and scatter)
while simultaneously fitting for $\beta$. VELMOD takes as inputs TF
parameters, an object's redshift and one of the observables
(velocity-width or apparent magnitude), and maximizes the probability
of observing one given the other. The strength of VELMOD analysis is
that it neither assumes a one to one mapping from redshift space to
real space (accounting for errors due to triple-valued regions), nor
does it require calibration of the TF relation prior to its
implementation.

Forward VELMOD uses the velocity-width to predict a galaxy's apparent
magnitude, whereas the inverse method uses the apparent magnitude to
predict the velocity-width parameter. The forward method is strongly
dependent on selection effects, and thus requires a well-modeled
selection function. The inverse method, however, is much less sensitive to  
selection effects due to sample selection's possible weak
dependence on velocity-width. As the selection function of SFI++ is
rather difficult to model accurately due to its being a compilation of
various surveys with a range of selection criteria, we will make use
of the inverse method in our analysis. This analysis assumes a TF
relation of the form $\eta^0(M) = -b_{\rm{inv}}^{-1}(M-a_{\rm{inv}})$,
where $\eta = \log_{\rm{10}}(W)-2.5$, $M = m-5\log(d\sbr{L}(r))$ is
the absolute magnitude, $d\sbr{L}$ is the luminosity distance, and
where $b_{\rm{inv}}$, $a_{\rm{inv}}$ and $\sigma_{\eta}$ are the
slope, intercept and rms scatter of the inverse relation,
respectively.

The conditional probability of observing a measured velocity-width of
a galaxy with an apparent magnitude, $m$, and an observed redshift,
$z$, is given by:
\begin{equation}
P(\eta | m,cz) = \frac{P(\eta, m,cz)}{\int_{-\infty}^{\infty}d\eta\: P(\eta,
m,cz)},
\label{Peta}
\end{equation}
\noindent where
\begin{equation}
P(\eta, m,cz) = \int_{\rm{0}}^{\infty} dr\:P(\eta,m|r) \: P(cz|r)\:
r^2 \:,\end{equation}
\begin{equation}
\begin{split}
& P(\eta,m|r) \propto \Phi(m-\mu(r))\times\\
& \;\;\;\;\;\;\;\; S(m,\eta,r)\exp\left(-\frac{[\eta -
\eta^0(m-\mu(r))]^2}{2\sigma^2_{\eta}}\right),
\end{split}
\end{equation}
\begin{equation}
P(cz|r) = \frac{1}{\sqrt{2\pi\sigma^2_v}}\exp\left(\frac{-[cz -
cz_{\rm{pred}}]^2}{2\sigma^2_v}\right),
\end{equation}
\begin{equation}
(1 + z_{\rm{pred}}) =  (1 + z\sbr{cos}(r))(1 + \beta^* u(r)/c),
\end{equation}
where $z\sbr{cos}(r)$ is related to the comoving distance $r\equiv
H_0R$ through Equation (\ref{z2dist}), $S(m,\eta,r)$ is the selection
function, $u$ is the radial predicted velocity scaled to $\beta^{*} = 1$, 
$\sigma_v$ is the
scatter in the actual velocity compared to the linear theory prediction, and where
$\mu(r) \equiv 5\log r$ is the distance modulus. 
We take $\sigma_{v} = 150 \kms$ based on the tests discussed in Appendix A.
We can then compute
the product of the conditional probability $P(\eta | m,cz)$ over all
galaxies and minimize the quantity
\begin{equation}
\mathcal{L}_{\rm{IV}} = -2\sum_{i} \ln P(\eta_{i} | m_{i},
cz_{i})
\end{equation}
for the parameters $\beta^*$, $a\sbr{inv}$, $b\sbr{inv}$, and the
three components of $\mathbf{V_{\rm{ext}}}$.

\subsubsection{Forward Likelihood}
We use a maximum likelihood method first described in \citet{PikHud05}
which was developed to compare peculiar velocities obtained through
SNe surveys while accounting for triple-valued regions. In addition to
constraining $\beta^*$ and the three components of
$\mathbf{V_{\rm{ext}}}$ this method can be used to constrain
$\widetilde{h}$, a nusiance parameter which permits a rescaling of published distances. The forward likelihood method maximizes the probability of a galaxy having
its observed redshift
\begin{equation}
P(cz) = \int_{\rm{0}}^{\infty}drP(cz|r)P(r)\:,
\end{equation}
\noindent where 
\begin{equation}
P(cz|r) = \frac{1}{\sqrt{2\pi\sigma^2_v}}\exp\left(\frac{-[cz
-cz_{\rm{pred}}]^2}{2\sigma^2_v}\right),
\end{equation}

\begin{equation}\label{zpred}
(1 + z_{\rm{pred}}) =  (1 + z\sbr{cos}(r,\widetilde{h}))(1 + \beta^* u(r)/c),
\end{equation}
\noindent where $z\sbr{cos}$ is given through the relation
\begin{equation}
\widetilde{h}r = cz\sbr{cos}\left(1 -
\frac{1+q\sbr{0}}{2}z\sbr{cos}\right),
\end{equation}
\noindent and where
\begin{equation}
P(r) \propto
\exp\left(-\frac{[r-d]^2}{2\sigma_{d}^2}\right)[1+\delta_{\rm{g}}^*(r)],
\end{equation}
$d$ is the distance as determined by the peculiar velocity survey, and
where $\sigma_{d}$ is the uncertainty is the measured distance.  The
product of $P(cz)$ for all objects is then computed, from which the
quantity $\mathcal{L}_{\rm{FL}} = -2\sum_{i} \ln P(cz_{i})$ is
minimized.

\subsubsection{$\chi^2$ Minimization}
In addition to the comparison method discussed above we also perform a
simple $\chi^2$ minimization procedure to determine the best value of
$\beta^*$ and $\mathbf{V_{\rm{ext}}}$. For this minimization procedure
we compare the observed redshift of the object with the sum of its
measured distance and predicted peculiar velocity at that distance,
\ie
\begin{equation}
\chi^2(\beta^*) = \sum_{i} \frac{(cz_{i} -
cz_{\rm{pred}})^2}{\sigma^2_{d_i} + \sigma^2_v} \:,
\end{equation}
where $z_{\rm{pred}}$ is given by Equation (\ref{zpred}). Note that
this expression does not account for the effects of density
inhomogeneities along the line of sight. The recovered value of \b\ is
affected by inhomogeneous Malmquist bias and results from this method
are thus expected to be biased high as a result. Nevertheless, the
$\chi^2$ statistic is useful to assess goodness-of-fit and so is
included here.

\subsection{Results}\label{results}
Key results obtained through velocity-velocity comparison methods
discussed in \S\ref{vv} are summarized in Table \ref{key_results}. To
determine the value of $\beta^*$ at which $\mathcal{L}_{\rm{FL}}$ and
$\chi^2$ are minimized, a cubic function was fit to resultant
data. Comparison of the TF relation constants ($a_{\rm{inv}}$,
$b_{\rm{inv}}$) obtained to those found by \citet{MasSprHay06} are
shown in Table \ref{TFconsts}. The best fit value for all parameters,
including $\beta^*$ and its errors were obtained from 500 bootstrap
samples of both 2M++ and the peculiar velocity datasets. Through
bootstrap analysis it was found that peculiar velocity datasets and
2M++ contributed essentially equal amounts to overall parameter
errors. The value obtained for the best fit residual bulk flow, ${\bf
  V_{\rm{ext}}}$, is given in Table \ref{DPresults} along with the
bulk flow (50 \hmpc\ Gaussian-weighted mean of the velocity field) and
the predicted velocity of the LG arising from 2M++. A comparison of
$\chi^2$ with and without the residual bulk flow results in a difference of
34 for the 3 degrees of freedom, the residual bulk flow model 
is thus preferred at the 5.1$\sigma$ level.

\begin{figure}
\centering
\includegraphics[width=\linewidth]{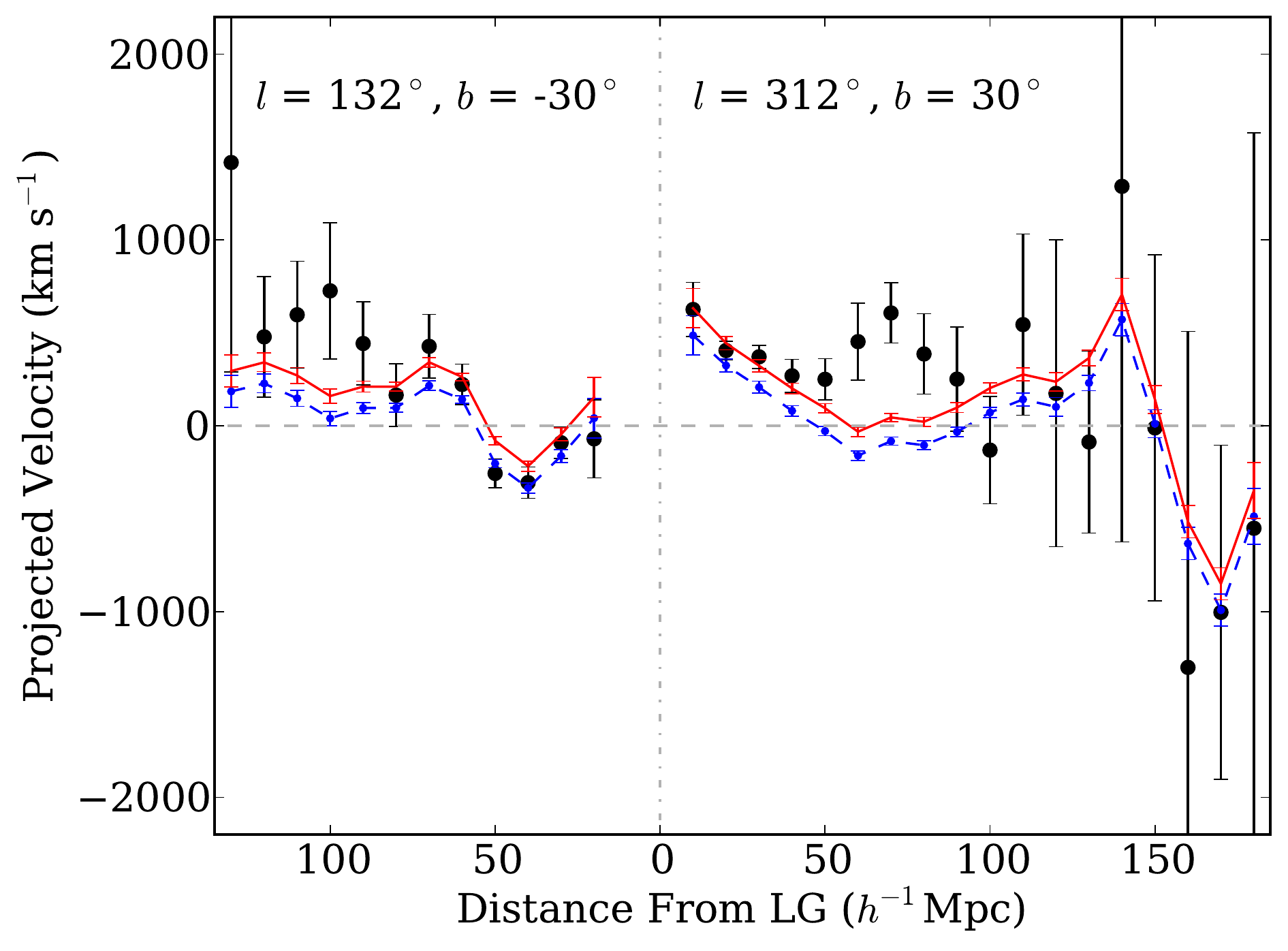}
\includegraphics[width=\linewidth]{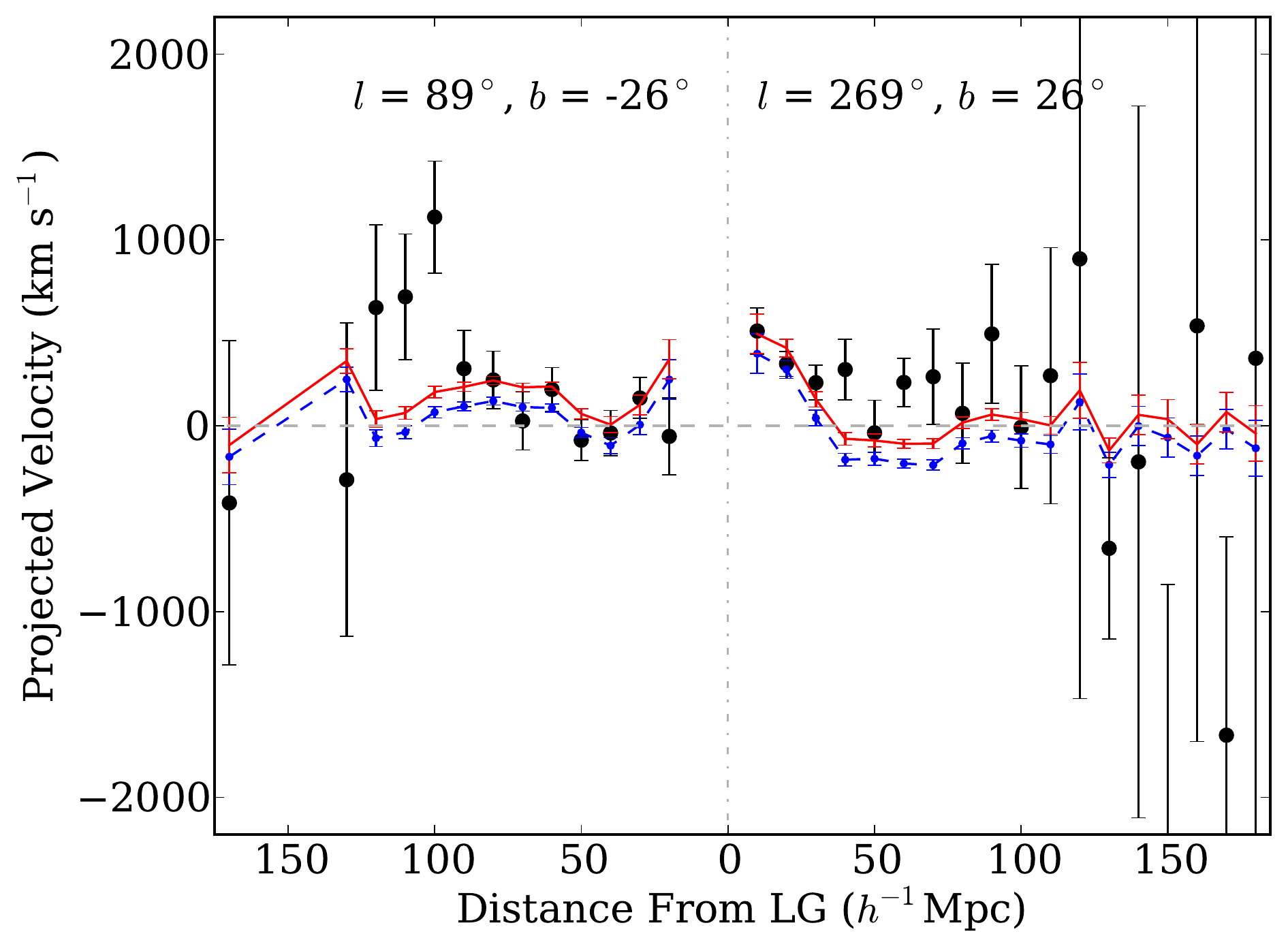}
\caption{Measured velocities of objects from A1 and SFI++ lying within
  a 30$^\circ$ cone of Shapley/Great Attractor (upper), and Hydra
  Superclusters (lower).  Velocities are projected on to the
  line-of-sight. Predictions arising from 2M++ for our best value of
  $\beta^*=0.43$ for these objects are shown above with and without
  the preferred residual bulk flow as connected blue dashed and red solid curves,
  respectively.  Values plotted are the error weighted mean of
  velocities within bins of 10 \hmpc. To the left of the origin the
  negative value of the peculiar velocity is plotted so that a bulk
  flow would appear as a constant offset.}
\label{LOSplots}
\end{figure}

For a qualitative illustration of the agreement of predicted
velocities arising from 2M++ with the published values from SFI++ and
A1 we have plotted the projected line-of-sight (LOS) velocity within a
$30^{\circ}$ cone centered on the Shapley Supercluster and Hydra
Supercluster in Figure \ref{LOSplots}. These two structures were
chosen as they lie in approximately the same direction as ${\bf
  V_{\rm{ext}}}$. From this figure it is apparent that predictions do
in fact follow the trends observed in measured
velocities. Furthermore, it is apparent that addition of ${\bf
  V}\sbr{ext}$ to predicted velocities seems to provide better
agreement, suggesting that such a residual bulk flow is in fact warranted.

In combination with a measurement of $\sigma^*_{\rm{8,g}}$, \b\ can be
used to constrain the cosmology dependent (and survey independent)
degenerate parameter combination
$f\sigma_{\rm{8}}=\beta^*\sigma^*_{\rm{8,g}}$. To measure
$\sigma^*_{\rm{8,g}}$ from 2M++ we follow a similar prescription to
that of \citet{EfsKaiSau90}, we compute $\sigma^*_{\rm{8,g}} =
\left<\sigma_{\rm{8,g}}(r)/\psi(r)\right>$ using counts in cells
within radial shells. Using this maximum likelihood scheme we obtain
the value $\sigma^*_{\rm{8,g}} = 0.99 \pm 0.04$, where the errors
quoted are derived from the scatter among different shells, as this
value was found to be more conservative then the formal errors from
the likelihood analysis. This value is in good agreement with that
found by \citet{Westover07} of $\sigma^*_{\rm{8,g}} = 0.98 \pm 0.07$
obtained by fitting projected correlation functions to 2MRS galaxies
within the magnitude range containing $L^*$ galaxies, \ie $-23.5 < K_s
< -23.0$. The product of the growth factor and non-linear
$\sigma\sbr{8}$ is thus $f\sigma\sbr{8} = 0.427 \pm 0.026$. By
adopting the value of $\Omega\sbr{m} = 0.3$, we can transform our
non-linear value of $\sigma\sbr{8}$ to a linearized value following
the prescription of \citet{JusFelFry10}. We in turn obtain the
constraint \fslin\ $= 0.401 \pm 0.024$. It is important to note that
linearization is only weakly dependent on the adopted value of
$\Omega\sbr{m}$ ($\Omega\sbr{m}=0.266$ results in \fslin\ $= 0.398 \pm
0.024$).  We compare these results with those obtained using
independent methods in the following section.
 
\begin{table}
\caption{Summary of best fit values of $\beta^*$ using different weighting
schemes, methods of analysis and peculiar velocity datasets. Results obtained
using luminosity weighting are indicated by (LW), whereas those obtained using
number weighting are indicated by (NW). Unless explicitly
indicated, all datasets were used for the method mentioned with the exception
of Inverse VELMOD which used all individual galaxies from SFI++.}
\centering
\begin{tabular}{l c c c}
\hline \hline
 & $\beta^*$ & $\chi^2/$(D.O.F.) \\
\hline
Forward Likelihood (LW) & & \\
A1 & 0.440 $\pm$ 0.023 & -\\
SFI++ Galaxy Groups & 0.429 $\pm$ 0.022 & -\\
SFI++ Field Galaxies & 0.423 $\pm$ 0.045 & -\\
{\bf All} & ${\bf 0.431 \pm 0.021}$ & -\\
\hline
Forward Likelihood (NW) & 0.439 $\pm$ 0.020 & -\\
\hline
Inverse VELMOD (LW) & 0.387 $\pm$ 0.048 & -\\
\hline
$\chi^2$ (LW) & 0.444 $\pm$ 0.026 & 2194/2899\\
\hline $\chi^2$ (NW) & 0.442 $\pm$ 0.028 & 2200/2899\\
\hline \end{tabular}
\label{key_results}
\end{table}

\begin{table}
  \caption{TF relation constants obtained through Inverse VELMOD analysis of
    SFI++ galaxies. Results listed are those obtained using a
    luminosity-weighting (LW) reconstruction scheme.}
\begin{tabular}{l r r}
\hline \hline
 & $a_{\rm{inv}}$ & $b\sbr{inv}$ \\
\hline
\citet{MasSprHay06} & -20.881 & -8.435 \\
This Study & -20.918 $\pm$ 0.012 & -8.19 $\pm$ 0.06 \\
\hline
\end{tabular}
\label{TFconsts}
\end{table}

\begin{table*}
  \caption{The bulk flow and motion of the LG arising from 2M++ for our best
    value of $\beta^*=0.43$. The best residual bulk flow, ${\bf V_{\rm{ext}}}$, that was
    fitted simultaneously with $\beta^*$ using the Forward Likelihood is also shown below.
    The bulk flow was computed by taking a 50 \hmpc\ Gaussian-weighted mean of the
    velocity field corresponding to $\beta^*=0.43$.}
\centering
\begin{tabular}{l c c c c c c}
\hline \hline
& $v_{\rm{x}}$ (\kms) & $v_{\rm{y}}$ (\kms) & $v_{\rm{z}}$ (\kms) &
$\|\vec{v}\|$ (\kms) & longitude ($^\circ$) & latitude ($^\circ$)\\
\hline
BF$\sbr{2M++}$ & -3 $\pm$ 8 & -72 $\pm$ 11 & 38 $\pm$ 11 & 81 $\pm$ 11 & 268
$\pm$ 6 & 28 $\pm$ 10\\

LG$_{\rm{2M++}}$ & -18 $\pm$ 27 & -422 $\pm$ 41 & 328 $\pm$ 37 & 535 $\pm$ 40 &
268 $\pm$ 4 & 38 $\pm$ 6 \\

${\bf V_{\rm{ext}}}$ & 89 $\pm$ 21 & -131 $\pm$ 23 & 17 $\pm$ 26 & 159 $\pm$
23 & 304 $\pm$ 11 & 6 $\pm$ 13\\

BF$\sbr{2M++}$ + \Vext & 86 $\pm$ 22 & -203 $\pm$ 26 & 55 $\pm$
28 & 227 $\pm$ 25 & 293 $\pm$ 8 & 14 $\pm$ 10\\

LG$_{\rm{2M++}}$ + \Vext & 71 $\pm$ 34 & -553 $\pm$ 47 & 345
$\pm$ 46 & 656 $\pm$ 47 & 277 $\pm$ 4 & 32 $\pm$ 6\\
\hline
\end{tabular}
\label{DPresults}
\end{table*}

\section{Discussion}
\label{discussion}
\subsection{Potential Systematic Effects}
In this section we discuss the various effects which could bias our
measurement of \b\ and our prediction of the motion of the
LG. Possible contributing factors include:
\begin{enumerate}
  \item Choice of smoothing length used for Gaussian kernel may skew our
  estimate of \b.
  \item Non-linear contributions to velocities from nearby small scale
  structure may systematically bias results.
  \item Triple valued regions may result in incorrectly reconstructed
  galaxy positions skewing velocity predictions derived therefrom.
  \item As the reconstruction is being done in the LG frame, at each iteration
  we are subtracting the motion of the LG from all galaxies, our procedure may
  thus be susceptible to the Kaiser rocket effect.
  \item Sparseness of survey may result in an under-representation of
  structures within the survey volume.
  \item Structures which lie in the ZoA are not included in 2M++ and
  could contribute to the direction and amplitude of the motion of the LG
  as well as influencing predictions of peculiar velocities of galaxies near
  the galactic plane.
  \item A non-linear relation between mass and luminosity as well as scatter
  in the underlying relation may influence predictions.
\end{enumerate}
\noindent We have addressed some of these concerns above, but will review
them again here for completeness.

In addressing (i), the width of the Gaussian kernel was chosen to be 4
\hmpc\ as this length was shown to be least biased when velocity
predictions were compared to those derived from
simulations. Furthermore, this smoothing length was found to produce
minimal scatter in the derived relation. We have accounted for (ii)
and by performing the full reconstruction and analysis on data derived
from N-body simulations. It was found that a value of $150$ \kms\
should be used for the scatter around predictions to account for these
effects. As for (iii), we used maximum likelihood methods which
integrate likelihoods along the line-of-sight.  As these methods
account for uncertainty in a galaxy's position, results should not be
sensitive to a misallocation of a galaxy in the event that it is lies
within a triple-valued region. The value of \b\ derived from this
analysis was found to be unbiased when performed on simulations. A
more complete discussion of the quantification of these systematics
through analysis of N-body simulations can be found in Appendix A.

As seen in Figure \ref{LG} although the direction of the LG is not
very susceptible to the value of \b, the overall amplitude varies by
$\sim$1200 \kms\ between \b$=0$ and \b$=0.8$. As we are doing the
reconstruction in the LG frame, an error in the estimate of the motion
of the LG may result in spurious distance estimates of
objects along this line of motion.  Putting objects at the incorrect
distance may in turn result in incorrect object weights, and in turn,
incorrect velocity predictions. This phenomenon has come to be known
as the Kaiser ``Rocket Effect'', as it was first discussed in
\citet{Kai87}. To account for this effect \citet{StrYahDav92} explored
a ``Kaiser Fix'' to the IRAS 1.2 Jy sample. This ``fix'' amounts to
altering the predicted distances of objects to
\begin{equation}
\begin{split}
&r = cz - \hat{{\bf r}}\cdot ({\bf V(r)} - {\bf
V(0)}\exp(-r^2/r_{\rm{K}}^{2})
\\
&\;\; - {\it {\bf V}}_{\rm{CMB}}[1 - \exp(-r^2/r^2_{\rm{K}})]),
\end{split}
\end{equation}
where ${\bf V}_{\rm{CMB}}$ is the velocity of the LG as inferred from
the CMB dipole, and where $r\sbr{K}$ is 1,000 \kms\ as determined by
the observed velocity correlation function \citep{BerGorDek90}. Note
that this fix assumes that galaxies more distant than $r_{\rm{K}}$ are
in fact at rest in the CMB frame, which may not be the case; indeed
our data suggests otherwise. Nevertheless, to estimate the sensitivity
of our results to this effect, we have implemented the Kaiser fix. We
find that the final estimate of \b\ differs by only 3\%, which is
small compared to the random errors.

The impact of survey sparseness on the methodology applied in this
work has been estimated in the past by \citet{PikHud05}. For the 2MASS
catalogue they found that under-sampling by 50\% produced negligible
results on their final estimate of \b\ (2-3\%). Furthermore, we have
accounted for the effects of sparse sampling by obtaining our quoted
results and errors from bootstrap resampling of both 2M++ and peculiar
velocity datasets. In addressing (vi) we measured \b\ at high
latitudes ($b>50^\circ$) and found no deviation from previous results
beyond that of the random errors.

When the density field was normalized to the same bias,
number-weighting and luminosity-weighting schemes yielded consistent
results. As the $\chi^2$ was found to be smaller for the luminosity
weighted result than that obtained with number weighting, we hereafter
will be quoting the best value of \b\ as that obtained from the
luminosity weighting scheme.  Results obtained in this paper account
for neither a non-linear relation between mass and luminosity, nor do
they account for scatter in mass-luminosity relation. More
sophisticated models have been proposed, such as the halo-model of
\citet{MarHud02}, and recently an iterative prescription to
reconstruct the density field from the distribution of halos
\citep{WanMoJin09}. We will consider implementation of such methods in
a future paper.

\subsection{Comparison With Other Results}
As the bias depends on luminosity and possibly morphology ({\it cf.} \citealt{PikHud05}), the value of
$\beta$ obtained is survey dependent. As 2M++ draws primarily from
2MASS, however, a loose comparison can be made with other values of
$\beta$ obtained therefrom. Most recently, in comparing the clustering
dipole of galaxies from 2MASS-XSC to predictions from linear theory
assuming a $\Lambda$CDM cosmology and convergence of the LG dipole
with that derived from the CMB, \citet{BilChoMam11} found $\beta=0.38
\pm 0.04$.  After constructing $\beta$ dependent predictions of
peculiar velocities, \citet{BraDavNus12} in turn estimate $\beta$ by
minimizing the scatter of predicted 2MASS absolute magnitudes about a
universal luminosity function and find $\beta=0.323 \pm
0.083$. \citet{DavNusMas11} expand the velocity field in spherical
harmonics and fit the inverse TF relation to SFI++ finding $\beta=0.33
\pm 0.04$. The TF data span a range of distances, and because the effective bias changes with distance it is difficult to compare this directly with our number-weighted result $\beta^{*} = 0.44\pm0.02$. The characteristic, or error-weighted,
depth of all individual SFI++ galaxies is $\sim 32 h^{-1}\hbox{Mpc}$. At this distance, the typical relative bias of a number-weighted sample with the magnitude limit of 2MRS is $\psi^N_{\rm{2MRS}} = b\sbr{eff}/b_{*} = 0.93$, and so applying this to the \citet{DavNusMas11} result yields $\beta^{*} = 0.31 \pm 0.04$

Our value of \fslin\ $= 0.40 \pm 0.02$ is in good agreement with those
obtained using the same methodology, such as that of
\citet{TurHudFel12} ($0.40 \pm 0.07$), that of \citet{PikHud05} ($0.44
\pm 0.06$), as well as with the weighted IRAS average of multiple
studies reported therein ($0.40 \pm 0.03$). As noted above, it is, 
however, in slight tension
with that found by \citet{DavNusMas11} ($0.31 \pm 0.04$). 
Note that where necessary the values quoted here have been linearized following the
procedure discussed above in \S\ref{results}.

\subsection{Cosmological Implications}

\subsubsection{The value of \fslin}
We can also compare our value \fslin\ to constraints placed on a
degenerate combination of $\Omega\sbr{m}$ and $\sigma_8$ through
independent means.  In particular our value is in excellent agreement
with a different peculiar velocity probe, 
namely measurements of $f(z)\sigma_8(z)$ at different redshifts via redshift
space distortions, which yield a best-fit value of \fsig\ $= 0.40 \pm
0.02$ \citep{HudTur12}.

\begin{figure}
\centering
\includegraphics[width=\linewidth]{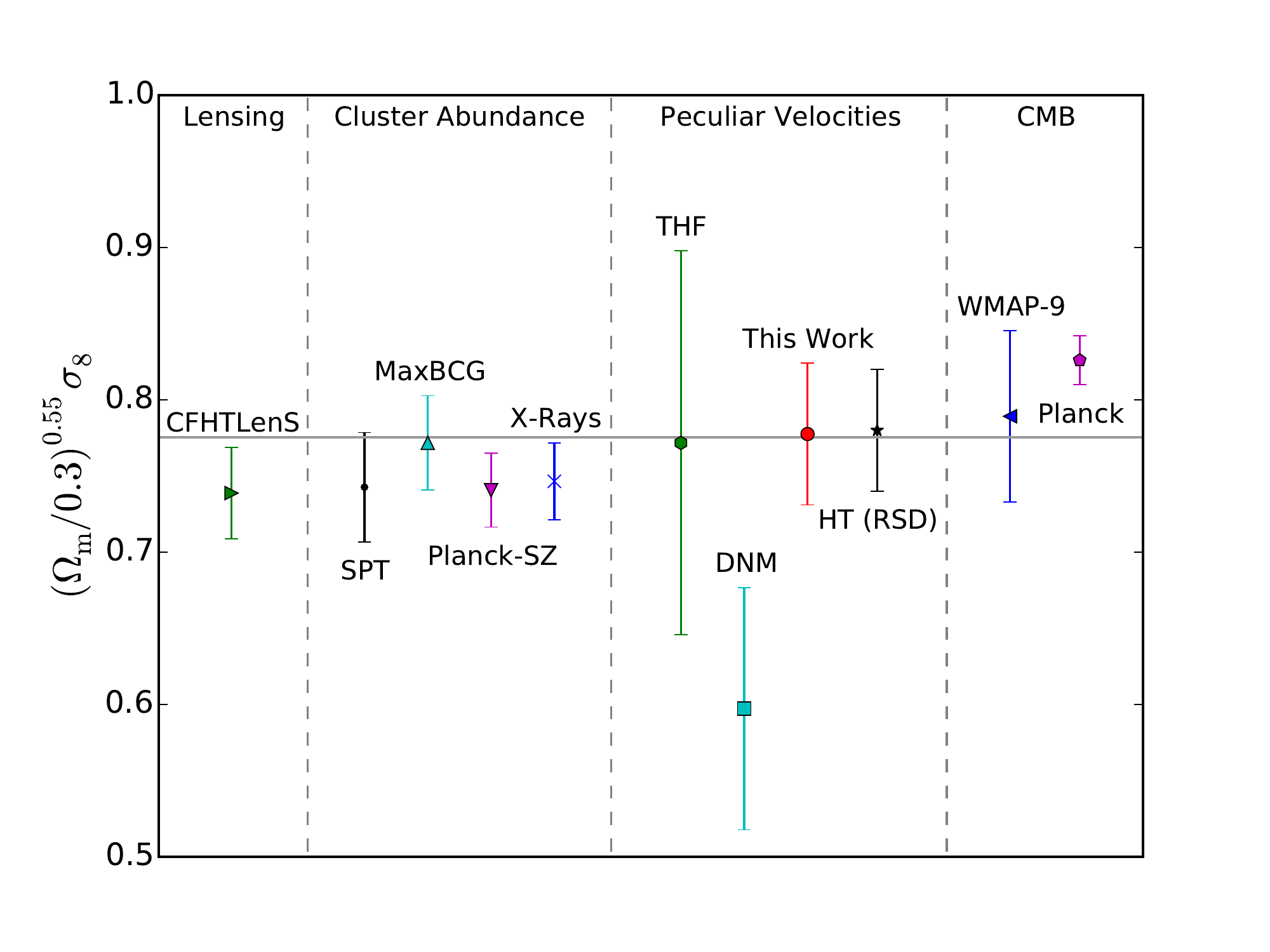}
\caption{Comparison of $f\sigma_{8, \rm{lin}}$ measured
  results. Values plotted above derived from weak-lensing
  \protect \citep[CFHTLenS]{KilFuHey13} and cluster abundances
  [\protect \citet[SPT]{ReiStaBle12}, \protect \citet[MaxBCG]{RozWecRyk10},
  \protect \citet[Planck-SZ]{PlanckXX}, \citet[X-rays]{VikKraBur09}] have
  assumed a value of $\Omega_{\rm{m}}=0.3$ in mapping constraints to
  $\Omega_{\rm{m}}^{0.55}\sigma_8$. Results obtained through previous analyses
  of measured peculiar velocities are also shown 
  [\protect \citet[THF]{TurHudFel12},
  \protect \citet[DNM]{DavNusMas11}], 
  as well as from redshift space distortions \protect \citep[HT]{HudTur12}.   
  CMB results are from WMAP9 and the Planck Collaboration (2015) \protect \nocite{Planck15XIII}.
  The horizontal line is the error-weighted mean of all values
  ($f\sigma\sbr{8} = 0.400 \pm 0.005$), shown here for reference.}
\label{fsigma8}
\end{figure}

An analysis of second and third-order
weak-lensing aperture-mass moments measured by CFHTLenS yields
$\sigma_8(\Omega_{\rm{m}}/0.27)^{0.6} = 0.79\pm0.03$
\citep{KilFuHey13}. Constraints can also be obtained from the number
counts and mass of galaxy clusters as measured through X-ray surface
brightness (\citealt{VikKraBur09}) and measurements of the
Sunyaev-Zeldovich (SZ) effect (\citealt{PlanckXX},
\citealt{ReiStaBle12}). Finally, we can obtain a value for $f\sigma_8$
from CMB temperature anisotropy from Planck \citep[\protect \OmSig\ $= 0.427
\pm 0.010$;][]{Planck15XIII} and WMAP9 \citep[\protect \OmSig\ $= 0.407 \pm
0.029$;][]{WMAP9}.  The measurements use different methods and are at
different redshifts, and so their dependence on $\Omega\sbr{m}$
differs in the exponent. To make a quantitative comparisons between
different results, we adopt $\Omega_{\rm{m}}=0.3$ (see Figure
\ref{fsigma8}). There is some tension between some results \eg\
\cite{KilFuHey13} and Planck-SZ \citep{PlanckXX} versus Planck CMB
temperature \citep{Planck15XIII}. The peculiar velocity result
presented here is consistent with all of these values.

\subsubsection{The motion of the LG}
For our best value of \b\ we can compare the predicted growth of the
LG velocity amplitude with the result that one would expect to measure
using linear theory for a $\Lambda$CDM cosmology (conditional on
$V_{\rm{CMB}}$). For \b$=$ \bestb, as determined from peculiar
velocity comparisons, we obtain the prediction for the motion of the
Local Group arising from 2M++ to be 535 $\pm$ 40 \kms\ in the
direction \lberr{268}{4}{38}{6}, only $10^\circ$ out of alignment with
the direction of the motion as inferred from the CMB dipole.  The
residual LG motion is therefore 100 $\pm$ 37 \kms\ in the direction
\lberr{303}{36}{34}{36}. This value is in reasonable agreement with
the best fit residual bulk flow obtained through peculiar velocity comparisons
in the CMB frame of 159 $\pm$ 23 \kms\ in the direction
\lberr{304}{11}{6}{13}. Inclusion of this residual bulk flow with the predicted
motion of the LG arising from 2M++ results in a total predicted motion
of 656 $\pm$ 47 \kms\ in the direction \lberr{277}{4}{32}{6}, in even
better agreement with both the amplitude and direction of the motion
as inferred from the temperature dipole of the CMB.

\begin{figure}
\centering
\includegraphics[width=\linewidth]{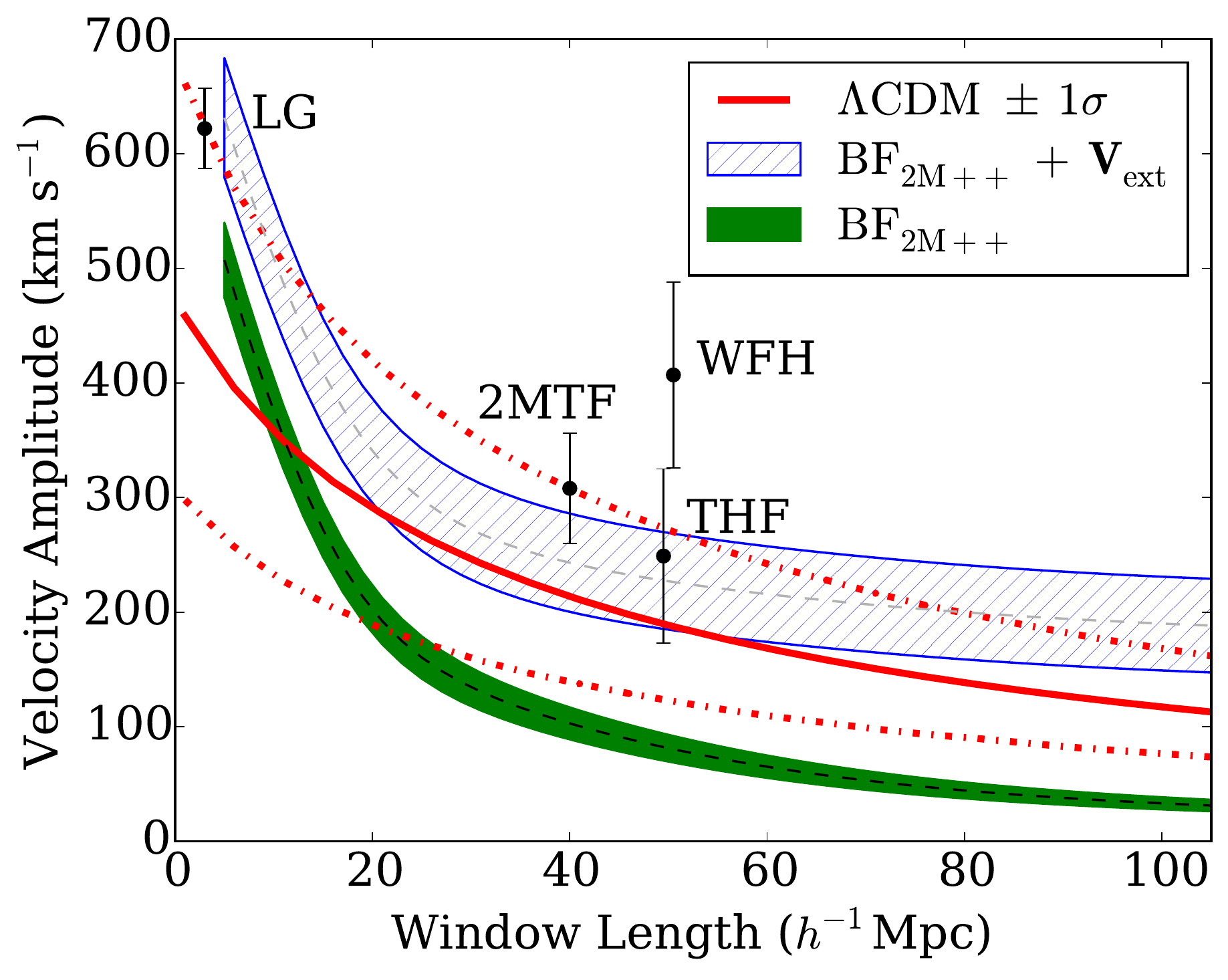}
\caption{Volume-weighted mean of predicted velocity field for Gaussian
  window of increasing scale centered on the Local Group. The inferred
  values from 2M++ with and without the residual bulk flow are shown by
  the dashed grey line with 68\% uncertainties in blue hatch, and a dashed
  black line, with uncertainties in solid green, respectively.  The
  predicted root-mean-square velocity for a $\Lambda$CDM WMAP9
  cosmology is shown as the red solid line, the cosmic scatter in the
  velocity amplitude distribution are shown as red dot-dash lines.
  Bulk flows in Gaussian-weighted spheres of radius 
  40 \hmpc\ and 50 \hmpc\ are shown for the results of 
  \protect \citet[2MTF]{HonSprSta14}, 
  \protect \citet[THF]{TurHudFel12} and  \protect \citet[WFH]{WatFelHud09} . 
  The LG motion is also shown, plotted at a radius of 3 \hmpc.  }
\label{BF}
\end{figure}

\subsubsection{The residual bulk flow}

We find that the amplitudes and directions of \Vext\ fit to each of the 
SFI++ and A1 SNe datasets separately are consistent with one another. 
Furthermore, comparing A1 with PSCz (of comparable depth to 2M++), 
\citet{TurHudFel12} found a residual flow of 
$V\sbr{x}=144\pm 44$ \kms, $V\sbr{y}=-38\pm
51$ \kms, $V\sbr{z}=20\pm 35$ \kms, in reasonable agreement with the
values found here of $V\sbr{x}=89\pm 21$ \kms, $V\sbr{y}=-131\pm 23$
\kms, $V\sbr{z}=17\pm 26$ \kms. This suggests that the residual bulk flow is not
an artifact of either the analysis or redshift-catalogue and is
sourced by structures outside the 2M++ and PSCz volumes.

We can also use the 2M++ density field to predict the BF and compare this to 
the BF expected in a $\Lambda$CDM universe in Figure \ref{BF}. We have
plotted this comparison for the Gaussian-weighted mean of the 2M++
velocity field. It is apparent from this figure that the resulting
bulk flow from our analysis is in agreement with that expected for a
$\Lambda$CDM universe.  Combining the cosmic variance in quadrature
with observational errors, comparison of the measured bulk flow of a
100 \hmpc\ Gaussian with predictions from $\Lambda$CDM yield a
$\chi^2$ of 1.4 for 3 degrees of freedom; clearly the measured value
agrees well with the predicted value from the standard cosmological
model.

\subsubsection{A large-scale underdensity?}

There have been recent claims that the Local Universe ($\sim$150--200
\hmpc) is under-dense (\citealt{WhiSha13},
\citealt{KeeBarCow13}). Such a phenomenon might account for the
discrepancy between the larger value for the Hubble parameter when
measured locally ($z \approx 0$) and that obtained from studies of the
CMB temperature anisotropies.

\begin{figure}
\centering
\includegraphics[width=\linewidth]{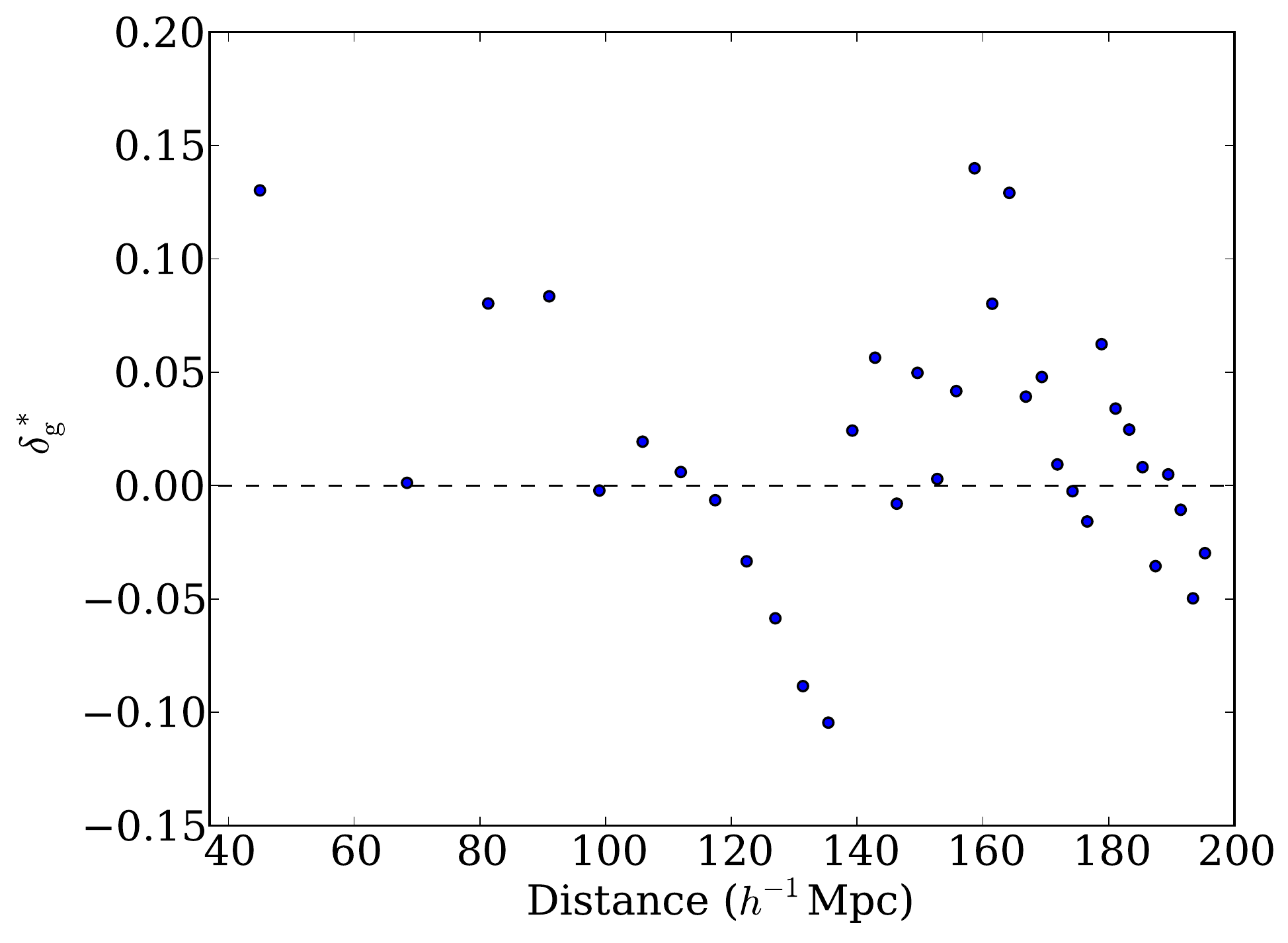}
\caption{The mean density contrast of concentric spherical shells of
  equal volume centered on the Local Group.}
\label{DensityShells}
\end{figure}

Although the majority of 2M++ lies within the suggested underdensity,
we have nonetheless explored the possibility of a under-dense volume
{\it within} 2M++. The luminosity-weighted density contrast of 2M++ in
shells is shown in Figure \ref{DensityShells}. We have not observed
any global systematic rise in density towards the periphery of the
survey.

To compare our results with others in more detail, note that 
\citet{WhiSha13} use redshift data from three large regions: 
6dF-SGC, 6dF-NGC \& SDSS-NGC.
Within $z < 0.05$, they quote 
mean density contrast of $\bar{\delta}\sbr{g} = -0.40\pm0.05$,
$0.04\pm0.10$ and $-0.14\pm0.05$, respectively.  For the same $z <
0.05$ volumes, we find
density contrasts of $\bar{\delta}_{*} = -0.17$, $0.01$ and $0.03$
respectively, where the density is normalized with respect to the mean
density within 200 \hmpc\ ($z\sim 0.067$).  \cite{BoeChoBri14} studied
the large-scale densities of X-ray clusters.  For the 6dF-SGC and
6dF-NGC regions, they find
mean \emph{cluster} density contrast of $\bar{\delta}\sbr{cl} = -0.55\pm0.10$ and
$0.02\pm0.17$ within $z < 0.05$.
However, as they point out, galaxy clusters are highly biased
($b\sbr{cl} \sim 2.7$) and so the corresponding mean matter density
contrasts are $\bar{\delta}\sbr{m} = -0.20\pm0.04$ and $0.01\pm0.06$.
These latter numbers are in good agreement with our nearly-unbiased
galaxy luminosity results. We conclude that, while the 6dF-SGC region
may be mildly underdense within $z \lesssim 0.05$, there is no evidence for a
large-scale void.

\subsubsection{Prospects for the future}
There are several upcoming peculiar velocity surveys which should
dramatically improve the constraints on both $\beta^*$ and
\fsig. Among these is the survey dubbed ``Transforming Astronomical Imaging surveys
through Polychromatic Analysis of Nebulae'' (TAIPAN). Using the
UK Schmidt telescope, it is estimated that TAIPAN will acquire
$\sim$45,000 Fundamental-Plane velocity measurements out to a redshift
of 0.2 (\citealt{KodBlaDav13}). The next generation of Tully-Fisher
(TF) peculiar velocity surveys include the Widefield ASKAP L-band
Legacy All-sky Blind surveY (WALLABY, \citealt{KorSta09}), and the
Westerbork Northern Sky HI Survey (WNSHS)\footnote{http://www.astron.nl/$\sim$jozsa/wnshs/}. An HI survey acquired using
the Australian Square Kilometer Array Pathfinder (ASKAP), WALLABY is
planned to cover 3$\pi$ steradian of sky. Its Northern Hemisphere
counterpart, WNSHS, is planned to cover remaining $\pi$ steradian of
the sky using the Westerbork Synthesis Radio. It is estimated that
these surveys will obtain a total of $\sim$32,000 velocity
measurements, and along with TAIPAN will not only enable $k$-dependent
measurements of \fsig\ but will improve constraints on this parameter
combination at low-redshift ($z\leq0.05$) to within 3\%
(\citealt{KodBlaDav13}). Clearly constraints on cosmology through
peculiar velocities has a very promising future.

\section{Conclusion}
\label{conclusion}
Under the assumption of gravitational instability, we reconstructed the 
galaxy density field within 200 \hmpc\
using 2M++, a composite all-sky redshift redshift survey with high
completeness. We compared the predicted peculiar velocities arising from this
density field to the measured peculiar velocities of SFI++ and the
First Amendment supernovae dataset using various comparison methods in
order to measure \b$=\Omega_{\rm{m}}^{0.55}/b^*$. Using the VELMOD
maximum likelihood comparison method with SFI++ spiral galaxies we
simultaneously solved for the inverse TF relation zeropoint and slope,
\b, as well as a residual bulk flow due to sources external to the 2M++ volume, denoted \Vext. We similarly compared
peculiar velocities datasets to predictions from 2M++ using a
forward-likelihood method in order to constrain \b\ and \Vext. All
methods and data subsets yielded consistent values of \b, with our
final result being \b$=f(\Omega_{\rm{m}})/b^*=$\bestb. Combining our
value of \b\ with $\sigma^*_{\rm{8,gal}} = 0.99 \pm 0.04$ as measured
from 2M++ for \b\ $=$ 0.43, we in turn measured the parameter
combination $f\sigma\sbr{8,lin}$ $=$ \bestfsig. This value was found
to be consistent with the majority of results obtained by independent
means, including those of WMAP9 and Planck.

For our measured value of \b\ we computed the velocity of the Local
Group as predicted by linear theory arising from the reconstructed
density field. Our value for the velocity was found to be consistent
with the theoretical value that would be measured for a survey of this
depth in a $\Lambda$CDM Universe. Combining our predicted value for
the motion of LG arising from 2M++ for our best value of \b\ with the
value of \Vext\ obtained through comparing predicted velocities with
peculiar-velocity surveys, we predict a motion of the LG to be $660
\pm 50$ \kms, towards \lberr{277}{4}{32}{6}, only 5$^\circ$ out of
alignment with the direction as inferred from the CMB
dipole. Similarly, with addition of this residual bulk flow to the 50 \hmpc\
Gaussian-weighted mean of the velocity field, we obtained a predicted
bulk-flow of $230 \pm 30$ \kms\ towards \lberr{293}{8}{14}{10}, an
amplitude that is consistent with that expected for a $\Lambda$CDM
Universe.  We note, however, that although we find the inclusion of
\Vext\ is preferred at the 5.1$\sigma$ level, it is unclear whether
this residual bulk flow merely compensates for imperfect mapping 
between luminosity and mass towards the periphery of 2M++, or whether it is due to
structures outside the 2M++ volume.  Future work using more sophisticated
biasing schemes may help in answering this question. 

The resulting 2M++ density and peculiar velocity fields obtained from this analysis
are made available at 
cosmicflows.uwaterloo.ca
and 
cosmicflows.iap.fr
.

\section{Acknowledgements}
All authors acknowledge support from the Natural Sciences and
Engineering Council of Canada. This work made in the ILP LABEX (under
reference ANR-10-LABX-63) was supported by French state funds managed
by the ANR within the Investissements d'Avenir programme under
reference ANR-11-IDEX-0004-02.

\bibliographystyle{mn2e}

\appendix

\section{Tests of the Reconstruction with N-body Simulations}\label{apdx}

In this Appendix, we focus on two possible sources of systematic bias
in the reconstructed density and velocity fields.  First, when
constructing the density field from a set of point density tracers, it
is necessary to smooth to obtain a continuous density field. If the
smoothing length is too short, density contrasts are high and linear
theory is no longer applicable.  If the smoothing length is too long,
then the density contrast is suppressed and velocities are
underpredicted. Second, in ``reconstructing'' real-space positions
from redshift-space, the iterative technique discussed in \S3 may also
introduce a systematic bias in the recovered density field and hence
in the fitted value of $\beta$.

\subsection{Effect of smoothing}\label{subsec:Smoothingbias}

\cite{BerNarWei00} used N-body simulations to show that when predicted
velocities derived from smooth density fields are compared to measured
(unsmoothed) velocities, the recovered value of $\beta$ depends on the
smoothing. For Gaussian smoothing, unbiased results were obtained for
a smoothing radius between 4 and 5 \hmpc. It is interesting to confirm
these results.

Here we use an N-body simulation of $512^{3}$ particles in a 500
\hmpc\ periodic box using GADGET-2 (\citealt{Spr05}). The cosmological
parameters of this simulation are as follows: $\Omega_{m} = 0.266$ ,
$\Omega_{\Lambda} = 0.734$, $h = 0.71$, and where each particle is
$6.83 \times 10^{10}$ $h^{-1} M_{\odot}$. From the particle positions
and velocities of the simulation a halo catalogue was formed using
ROCKSTAR (\citealt{BehWecWe13}) consisting of 693948 halos between
$5.5 \times 10^{11}$ and $2.2 \times 10^{15}$ $h^{-1}$ $M_{\odot}$
(between 8 and 31809 particles).  Either halo or particle positions
were smoothed to create a smooth halo or particle density field.

In \figref{SlopeBySmooth} we plot both the slope of regression between
smoothed predicted and N-body velocities, as well as the scatter about
this relation, both as a function of smoothing scale.

For particle velocities compared with linear theory predictions from
the smoothed particle density field, the slope is unbiased at a
smoothing scale of $\sim$5 \hmpc.  The scatter is $\sim$250 \kms.
The results are similar for halos, but the scatter is significantly
lower: $\sim$150 \kms.  This is because the particle velocity field is
a sum of the motion of the halos themselves plus the internal motion
of particles with respect to the halos.  In both cases, the scatter is
minimum for a smoothing scale of $\sim$4 \hmpc.

\subsection{Reconstructing the Halo Distances from Redshifts}
With actual galaxy data, the true distances are unknown but redshifts
are available. In this paper, we use an iterative reconstruction
method to map a galaxy's position from redshift space to real space.
We will refer to these reconstructed coordinates as ``recon-space.''
In the N-body simulation, we emulate this by placing halos in
redshift-space, and then iteratively reconstruct the density field,
slowly increasing $\beta$.

\begin{figure*}
\centering
\includegraphics[width=0.5\textwidth]{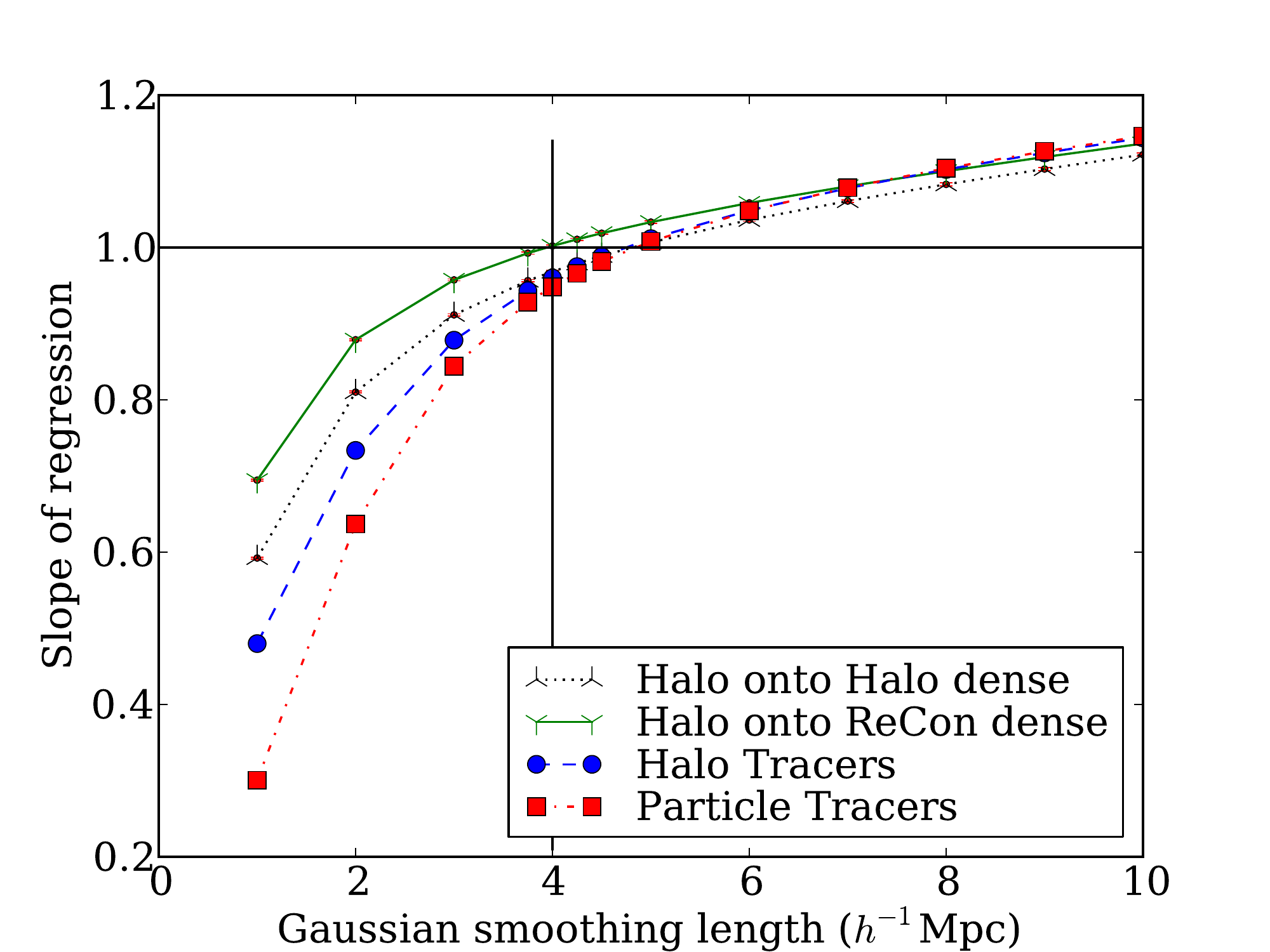}\includegraphics[width=0.5\textwidth]{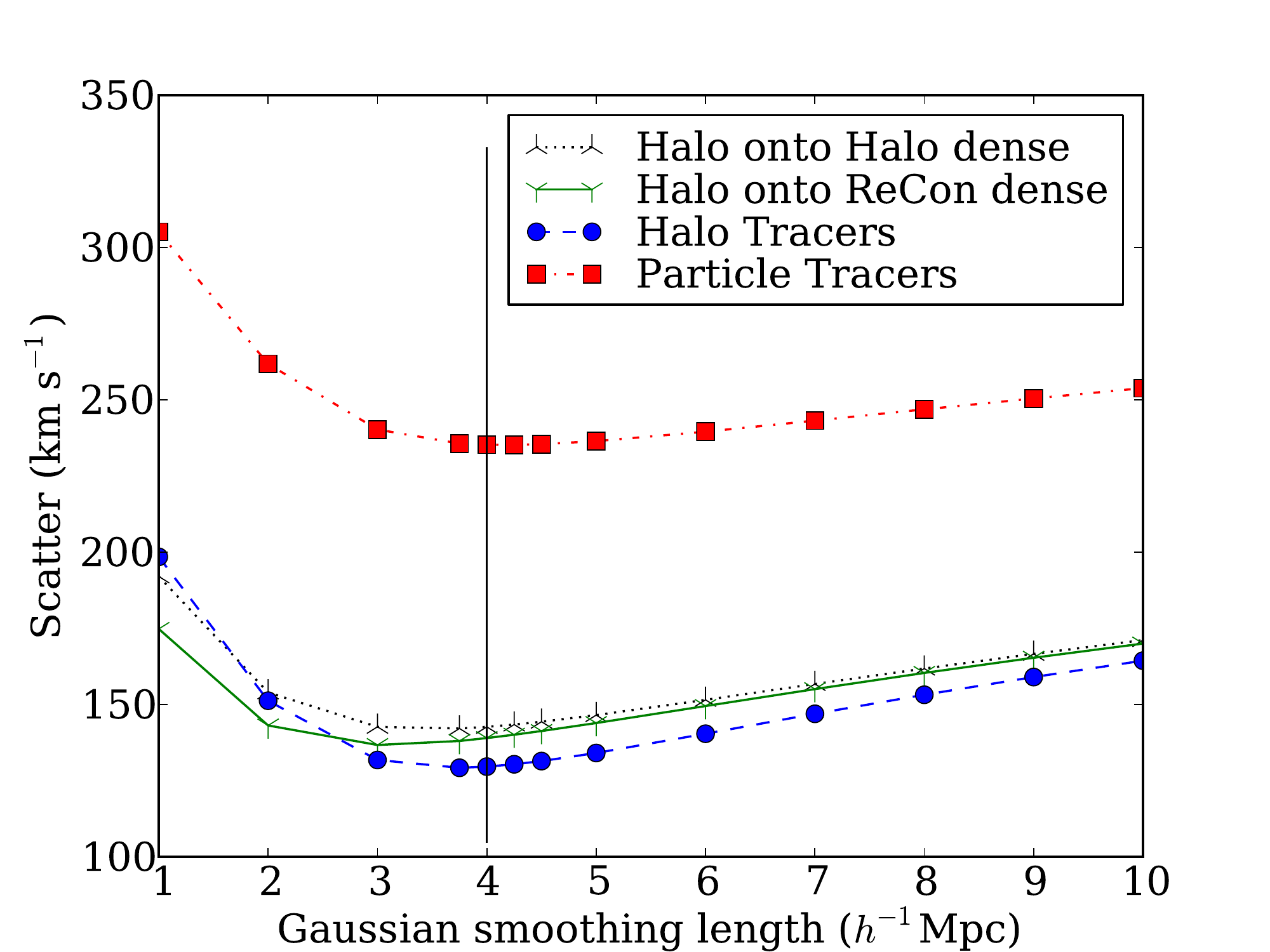}
\caption{The fitted slope of the regression between N-body observed
  peculiar velocities and the linear theory predictions as a function
  of smoothing scale is shown in the left panel. In the right panel
  the scatter about the same regression is plotted.  Either particles
  or halos can be used as tracers of the density field, or as tracers
  of the velocity field. The blue circles and red squares represent
  halo tracers and particle tracers, respectively.  The black curve
  with downwards pointing `Y' show the resulting scatter when the
  known halo velocities are compared to the predictions from the halo
  density field. The green curve with `Y'-shaped symbols show the
  resulting scatter when the known halo velocities are compared to the
  predictions from the reconstructed halo density field. Note that the
  reconstruction process shows very nearly unbiased results when the
  field is smoothed with a Gaussian kernel that is 4 \hmpc\ in
  length.}
\label{fig:SlopeBySmooth}
\end{figure*}

\begin{figure*}
\centering
\includegraphics[width=0.5\textwidth]{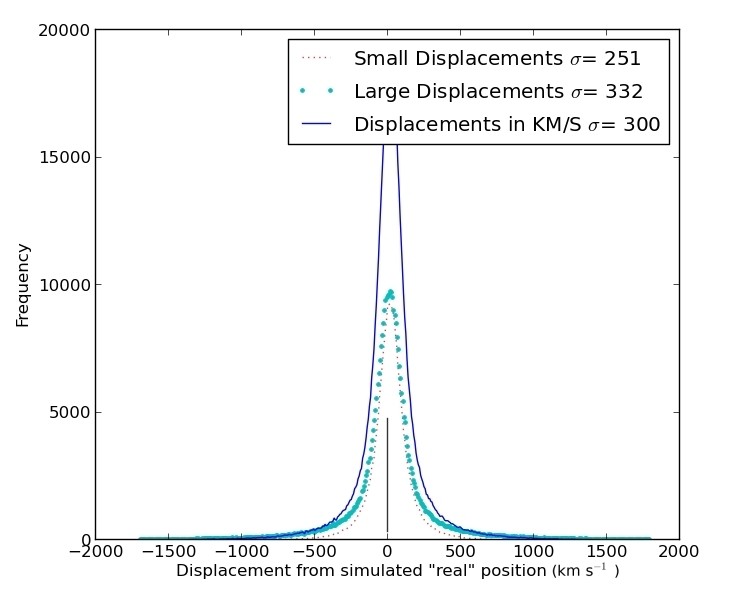}\includegraphics[width=0.5\textwidth]{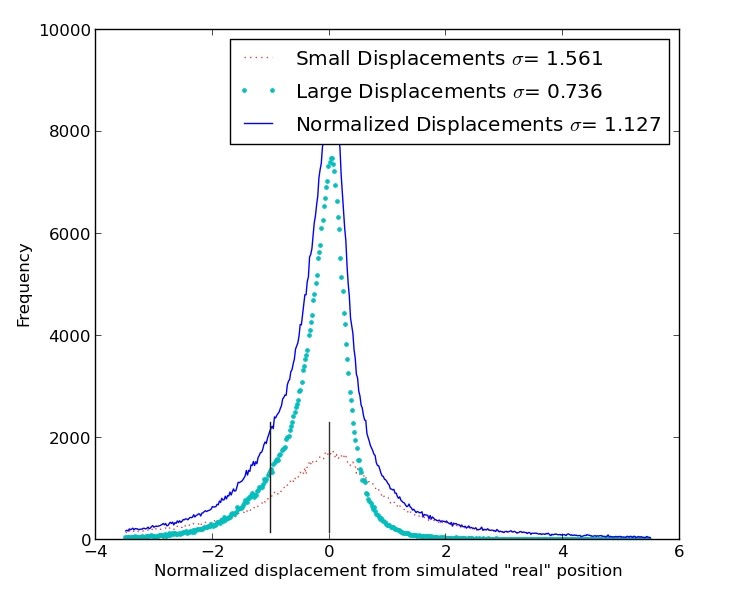}
\caption{Two sample histograms showing the relative displacement of
  halos in recon-space to their real-space positions. The recon-space
  positions used are those which correspond to the number of
  iterations required to recover the known value of $\beta$. The left
  panel illustrates the absolute displacement in \kms, whereas the
  right panel shows the displacement of halos normalized by the
  difference between real-space and redshift-space positions. As such,
  a normalized displacement of -1 corresponds to a halo lying at its
  redshift-space position, and a normalized displacement of 0
  corresponds to a halo returning to its real-space position. The
  solid blue curve corresponds to all halos, the turquoise dotted
  curve corresponds to halos with known velocities greater than 150
  \kms, and the dotted red curve corresponds to halos with known
  velocities less than 150 \kms. Halos which have a normalized
  displacement less than -1 have moved away from their initial
  positions due to lying within triple-valued regions. Halos which
  have a normalized displacement greater than 0 have moved to their
  initial positions and beyond.}
\label{fig:ThreeDirections}
\end{figure*}

\figref{ThreeDirections} shows the difference between real-space
positions and recon-space positions.
The left panel illustrates the absolute displacement in units of km
s$^{-1}$.
From this panel, it is apparent that in recon-space the majority of
halos do in fact return to a location close to their real-space
positions, with an error that is typically $\sim 300$ \kms.  This is
smaller than the smoothing scale of $\sim 4 \hmpc$. However, in
practice this reconstruction error will act as an additional source of
smoothing. We will return to this point below.

The right panel shows the displacement of each halo from its
real-space position normalized by the difference between the
real-space and redshift-space positions (i.e. by their peculiar
velocities).  In this way the halos which lie in redshift-space have a
normalized displacement of $-1$, and those which have returned to
real-space have a normalized displacement of $0$. With each successive
iteration tracers would all ideally return to $0$, {\it i.e.} return
to the position they had in real-space. Halos which have a normalized
displacement more negative than $-1$ have moved away from their
initial positions. We argue below (\S \ref{TVR}) that these are halos
in triple-valued regions.
From this figure it can be seen that the typical tracer does in fact
return to its real-space position. However, the distinction between
low and high peculiar velocity tracers is more pronounced. The tracers
with a low peculiar velocity have a symmetric distribution of
reconstruction error.  The halos which initially have a high velocity,
however, have a more skewed distribution which likely results in part
from triple-valued regions.

Having reconstructed the density field in recon-space, we return to
the question of how well this field, once smoothed, can predict
peculiar velocities.  \figref{SlopeBySmooth} shows the results of this
comparison.  The fitted slope obtained from the reconstructed density
field is systematically offset in comparison to those obtained from
the real-space halo density field.  The key result is that predicted
velocities derived from the reconstructed smoothed halo fields yields
an unbiased value of the slope, {\it i.e.} $\beta$, when a Gaussian
kernel of 4 $h^{-1}$Mpc is used. This value is smaller than the value
of $\sim$5 \hmpc\ found for halos in real-space, because the
recon-space density field is effectively pre-smoothed (in the redshift
direction) by the $\sim$ 300 \kms\ reconstruction error discussed
above.

\section{Iterative Reconstruction Near Triple-Valued Regions}
\label{TVR}
One problem with any reconstruction method is correcting for the fact
that mapping from redshift to position is multivalued, i.e. there can
be several distances which map to the same redshift.

\begin{figure}
\centering
\includegraphics[width=0.49\textwidth]{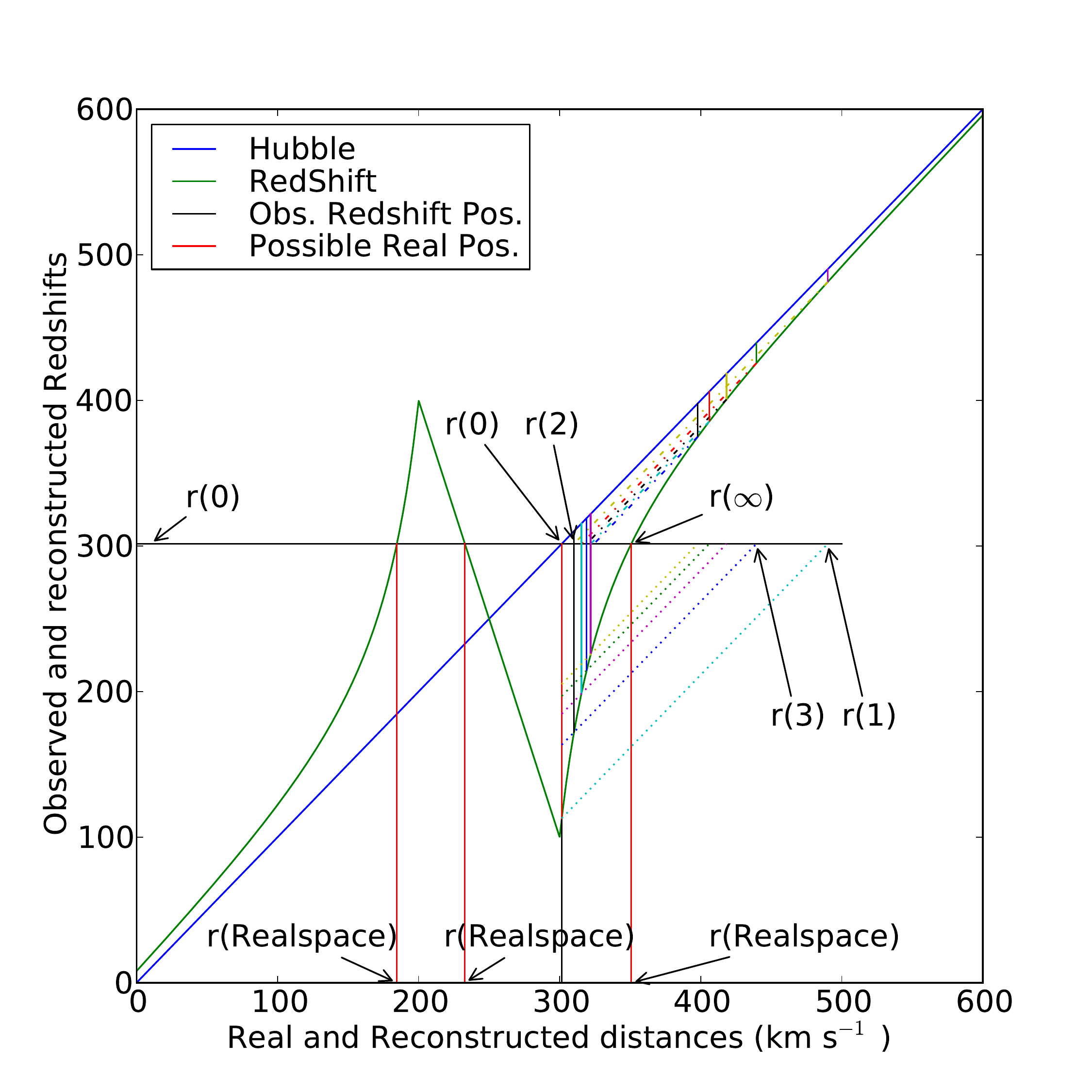}
\caption{Simplified reconstruction of one test object near a fixed
  potential well. In a Universe without structure, Hubble's law (blue
  solid line $y=x$) would be sufficient to convert from redshift to
  real-space positions. In a Universe containing a spherical top-hat
  overdensity, however, the relation is complicated by infall towards
  the well (green solid curve). An object with an observed redshift in
  excess of 300 \kms\ can be seen to have originated from three
  possible real-space positions (vertical red solid lines). The
  reconstruction process starts by placing the particle at it's
  observed redshift, predicting its velocity, and then subtracting
  that prediction from the original observation (the dotted lines
  parallel to the Hubble law line) to get a new predicted position. In
  most situations this new position will be beyond one of the possible
  real-space positions where the predicted infall velocities are
  small. Repeating the process here results in a much smaller velocity
  offset which can be projected (dot-dashed lines) back to the
  observed velocity to find the next reconstructed position. The
  iterative reconstruction converges in most situations to one of the
  two outer possible real-space positions. Reconstructed positions
  never converge, however, to the central possible real-space
  position.}
\label{fig:tvr_plot}
\end{figure}

To give an illustration as to how this affects the iterative
reconstruction method, we have created a simple toy model in
\figref{tvr_plot} which shows the velocity field around a spherical
overdensity. One can see that for any object with an observed redshift
in the range $\sim 100$ to $\sim 400 \kms$, there are multiple
distances that correspond to that given redshift.  As a concrete
example, consider a galaxy at a redshift of 300 \kms. We will consider
a simplified version of our reconstruction scheme in order to
illustrate the general behaviour. In this simplified version, the
value of $\beta$ is held fixed, i.e. it is not increased
adiabatically, so the amplitude of the peculiar velocity curve is
fixed and the curve $cz(r)$ is also fixed.

In the first step of this iterative process, we place the galaxy at a
distance corresponding to its observed redshift: 300 \kms\ at position
r(0) in \figref{tvr_plot}. At this location the predicted peculiar velocity 
is approximately $-190 \kms$, so the object is moved to its first reconstructed position
of $490 \kms = 300 \kms - (-190\kms)$ at position r(1). At this
position it has a predicted peculiar velocity of approximately $-20
\kms$, and so this predicted peculiar velocity is again subtracted
from its observed redshift of 300 \kms, resulting in a reconstructed
distance of 320 \kms indicated r(2). In this example, each successive
loop brings the reconstructed position closer to the outer redshift
solution of approximately 360 \kms\ indicated r($\infty$).  The same behaviour occurs for
redshifts between 250 and 400 \kms, the final position converges to
the third (outer) of the three values.
For redshifts between 100 and 250 \kms, in contrast, the recon
position converges to the first of the three values.  One can see that
the outer solutions are attractors, and the middle solution is
unstable.

The above toy model is for a test particle (e.g.\ galaxy) moving in a
fixed potential well (e.g.\ supercluster). It can also happen that two
similar-mass galaxies (or two clusters) may each respond to the
other's gravity. In this situation, the distances of the two objects
can ``leapfrog'', i.e.\ can exchange the order of their distances in
reconstructed space. This leads to a sign change of the predicted
peculiar velocity, which the leads to another ``leapfrog'' and a
subsequent sign change in predicted peculiar velocity. Hence some
objects show oscillatory behaviour in the reconstructed distance.

Note however that the simple model described above and in
\figref{tvr_plot} is an over-simplification of our reconstruction
process.  Of course, while in reality the density field used is a
smoothed Gaussian random sphere not a spherical top-hat overdensity,
qualitatively the behaviour around overdensities is similar.
More importantly, our iterative scheme is more complex than the one
outlined above.
First, we ``adiabatically'' increase \b\ at each
iteration. Consequently, the amplitude of the predicted peculiar
velocity field increase slowly and so the size of triple-valued
regions in redshifts space grows slowly.
Second, we reduce oscillatory behaviour by averaging the positions of
several successive iterations. Thus the convergence is better-behaved
than the toy model described above. But qualitatively, the effect
remains the same:
as a result of preferentially placing objects towards the periphery of
triple-value regions, the density field is effectively ``smoothed''
more in the regions of highest density.

\label{lastpage}
\end{document}